\documentclass[twocolumn]{aastex63}

\submitjournal{ApJ}

\shorttitle{High-frequency ALMA observations of $z\gtrsim6$ LBGs}
\shortauthors{Mitsuhashi et al.}

\graphicspath{{./}}
\renewcommand\baselinestretch{0.92}
\usepackage{amsmath}
\tabletypesize{\footnotesize}
\usepackage{multirow}
\usepackage{makecell}

\usepackage{ulem} 
\def\blue#1 {{\textcolor{blue}{#1}}\ }
\def\cii{$\text{[C\,{\sc ii}]}_{158}$}
\def\oiii{$\text{[O\,{\sc iii}]}_{88}$}
\def\oi{$\text{[O\,{\sc i}]}_{63}$}
\def\logRoiiicii{$\log (L_{\rm [OIII]88}/L_{\rm [CII]158})$}
\def\Roiiicii{$L_{\rm [OIII]88}/L_{\rm [CII]158}$}
\def\Rciioi{$L_{\rm [CII]158}/L_{\rm [OI]63}$}
\def\Roiiioi{$L_{\rm [OIII]88}/L_{\rm [OI]63}$}

\def\kms{\,km\,s$^{-1}$}

\def\red#1 {{\textcolor{red}{#1}}\ }

\begin{document}

\title{SERENADE III: Insight into the Origin of the High Dust Temperature and \\
High [O\,{\sc iii}]/[C\,{\sc ii}] Ratio at $z\gtrsim6$}

\correspondingauthor{Ikki Mitsuhashi}
\email{ikki0913astr@gmail.com}

\author[0000-0001-7300-9450]{Ikki Mitsuhashi}
\affiliation{Department for Astrophysical \& Planetary Science, University of Colorado, Boulder, CO 80309, USA}

\author{Yuichi Harikane}
\affiliation{Institute for Cosmic Ray Research, The University of Tokyo, 5-1-5 Kashiwanoha, Kashiwa, Chiba 277-8582, Japan}

\author{Hiddo S. B. Algera}
\affiliation{Institute of Astronomy and Astrophysics, Academia Sinica, 11F of Astronomy-Mathematics Building, No.1, Section 4, Roosevelt Rd, Taipei 106319, Taiwan, R.O.C}

\author{Tom J. L. C. Bakx}
\affiliation{Department of Space, Earth and Environment, Chalmers University of Technology, Gothenburg, Sweden}

\author{Andrea Ferrara}
\affiliation{Scuola Normale Superiore, Piazza dei Cavalieri 7, I-56126, Pisa, Italy}

\author{Akio K. Inoue}
\affiliation{Waseda Research Institute for Science and Engineering, Faculty of Science and Engineering, Waseda University, 3-4-1, Okubo, Shinjuku, Tokyo 169-8555, Japan}
\affiliation{Department of Physics, School of Advanced Science and Engineering, Faculty of Science and Engineering, Waseda University, 3-4-1, Okubo, Shinjuku, Tokyo 169-8555, Japan}

\author{Masatoshi Imanishi}
\affiliation{National Astronomical Observatory of Japan, 2-21-1 Osawa, Mitaka, Tokyo 181-8588, Japan}

\author{Kotaro Kohno}
\affiliation{Institute of Astronomy, Graduate School of Science, The University of Tokyo, 2-21-1 Osawa, Mitaka, Tokyo 181-0015, Japan}
\affiliation{Research Center for the Early Universe, Graduate School of Science, The University of Tokyo, 7-3-1 Hongo, Bunkyo-ku, Tokyo 113-0033, Japan}

\author{Yoshiaki Ono}
\affiliation{Institute for Cosmic Ray Research, The University of Tokyo, 5-1-5 Kashiwanoha, Kashiwa, Chiba 277-8582, Japan}

\author[0000-0001-6958-7856]{Yuma Sugahara}
\affiliation{Waseda Research Institute for Science and Engineering, Faculty of Science and Engineering, Waseda University, 3-4-1, Okubo, Shinjuku, Tokyo 169-8555, Japan}
\affiliation{Department of Physics, School of Advanced Science and Engineering, Faculty of Science and Engineering, Waseda University, 3-4-1, Okubo, Shinjuku, Tokyo 169-8555, Japan}

\author{Hideki Umehata}
\affiliation{Institute for Advanced Research, Nagoya University, Furocho, Chikusa, Nagoya 464-8602, Japan}
\affiliation{Department of Physics, Graduate School of Science, Nagoya University, Furocho, Chikusa, Nagoya 464-8602, Japan}
\affiliation{Cahill Center for Astronomy and Astrophysics, California Institute of Technology, MS 249-17, Pasadena, CA 91125, USA}

\author{Livia Vallini}
\affiliation{INAF-Osservatorio di Astrofisica e Scienza dello Spazio, via Gobetti 93/3, I-40129, Bologna, Italy}

\author[0000-0002-7051-1100]{Jorge A. Zavala}
\affiliation{National Astronomical Observatory of Japan, 2-21-1 Osawa, Mitaka, Tokyo 181-8588, Japan}

\begin{abstract}
We present an analysis of ALMA high-frequency observations of nine bright Lyman-break galaxies at $5.8<z_{\rm spec}<8.3$. 
Our sample consists of five galaxies at $z\sim6$ newly observed in Band-9 and/or 10, allowing us to better constrain the dust temperature ($T_{\rm dust}$) in a statistical sample of $z\gtrsim6$ galaxies. 
Our measurements of the dust temperature at $z\sim6$--9 suggest most of the star-forming galaxies show $T_{\rm dust}\sim40\,{\rm K}$ on average, whereas three galaxies show significantly higher $T_{\rm dust}$ ($\gtrsim60\,{\rm K}$). 
We find a potential negative correlation between $T_{\rm dust}$ and gas-phase metallicity with $T_{\rm dust}\propto Z^{-0.50\pm0.19}$, implying decreased dust shielding and efficient dust heating in low-metallicity environments.
Given the systematic offset of $T_{\rm dust}$ between $z\sim0$ and $z\sim6$--9 at fixed metallicity, we find $T_{\rm dust}$ is well described by sSFR as well as $Z$ with $\log T_{\rm dust}=-0.25^{+0.03}_{-0.04}\times(\log Z-0.57\times\log {\rm sSFR})+4.88^{+0.47}_{-0.43}$ across $z\sim0$ to $z\sim6$--9.
Simultaneously with the dust continuum, these observations cover the $\text{[O\,{\sc i}]}\,{63}\mu{\rm m}$ emission line for five galaxies, which traces dense neutral gas.
We find a lower \cii/\oi\ ratio in $z\gtrsim6$ galaxies than in local samples, suggesting that \cii\ becomes fainter due to collisional de-excitation at high gas density.
Combining the \oi\, \oiii, and \cii\ lines and \texttt{cloudy} modeling, our results imply $\log U_{\rm ion}\sim-2$ and $\log n_{\rm H}\,[{\rm cm}^{-3}]\sim2.5$ in $z\gtrsim6$ galaxies, which is $\sim3$--$10\times$ higher $U_{\rm ion}$ and $\sim2$--$3\times$ higher $n_{\rm H}$ than the local samples.
The combination of these enhanced $U_{\rm ion}$ and $n_{\rm H}$ naturally explains the high \oiii/\cii\ ratio at $z\gtrsim6$.
\end{abstract}


%
%
%
%
%
%
\section{Introduction}

Studies of galaxies at the epoch of reionization (EoR; $z \gtrsim 6$) are critical for understanding cosmic reionization and star formation activity in the early Universe. Observations with the Hubble Space Telescope (HST), various ground-based optical/near-infrared (NIR) facilities, and, more recently, the James Webb Space Telescope (JWST) have enabled detailed studies of EoR galaxies through rest-frame ultraviolet (UV) and optical wavelengths. These observations primarily trace star formation through UV photons emitted by young, massive stars \citep{1996ApJ...460L...1L,1996MNRAS.283.1388M,1999ApJ...519....1S,2007ApJ...670..928B,2010ApJ...725L.150O,2013ApJ...773...75O,2013ApJ...762...32C,2013ApJ...763L...7E,2015ApJ...803...34B,2015ApJ...810...71F,2020MNRAS.494.1894M,2023MNRAS.523.1009B,2023MNRAS.523.1036B,2023ApJS..265....5H,2024ApJ...960...56H,2024Natur.633..318C,2024ApJ...972..143C,2026OJAp....956033N} and the ionized interstellar medium \citep[ISM,][]{2002ApJS..142...35K,2006ApJ...644..813E,2008A&A...488..463M,2014MNRAS.442..900N,2015MNRAS.454.1393S,2017MNRAS.464..469S,2021ApJ...914...19S,2023ApJ...956..139I,2023ApJS..269...33N,2024A&A...684A..75C,2025NatAs...9..155Z,2025arXiv250810099S}.
In contrast, UV photons from young massive stars are easily absorbed by dust, and it is challenging to constrain the physical conditions of the neutral and dense gas phases at these redshifts. 
Such components are nicely traced at rest-frame far-infrared (FIR) wavelengths, where both dust thermal emission and key cooling lines originate (see \citealp{2020RSOS....700556H} for a review). 
The Atacama Large Millimeter/submillimeter Array (ALMA) provides a complementary view by probing dust-obscured star formation and the neutral and dense ISM through its high-sensitivity FIR continuum and line observations at high-$z$ \citep{1997ApJ...490L...5S,2006ApJ...645L..97I,2013Natur.496..329R,2013ApJ...774...68D,2017A&A...605A..42C,2022ApJ...931..160B}.

The cosmic dust-obscured star formation rate density (SFRD) has been extensively studied with FIR to submillimeter observatories and is known to dominate at $z \sim 0$--3 \citep[e.g.,][]{2001ApJ...556..562C,2009A&A...496...57M,2011A&A...528A..35M,2013A&A...553A.132M,2014ARA&A..52..415M}. 
Although the contribution of dust-obscured star formation at $z>4$ remains uncertain \citep{2021ApJ...923..215C,2021ApJ...909..165Z}, several studies suggest that it may still play a significant role \citep[e.g.,][]{2021Natur.597..489F,2023MNRAS.518.6142A,2023arXiv230301658F,2025ApJ...980...12S}. 
Accurately determining dust-obscured star formation rates requires robust constraints on the FIR spectral energy distributions (SEDs) of high-redshift galaxies.
The dust temperature ($T_{\rm dust}$) is a key parameter that characterizes the shape of the FIR SED \citep[e.g.,][]{2003MNRAS.338..733B,2012MNRAS.425.3094C} and thus directly impacts the estimation of the dust-obscured SFR. 
While $T_{\rm dust}$ is constrained at $z\lesssim4$ owing to extensive effort of FIR telescopes such as {\it Spitzer} and {\it Herschel} \citep{2014A&A...561A..86M,2015A&A...573A.113B,2018A&A...609A..30S,2020MNRAS.498.4192F,2022MNRAS.516L..30V,2026arXiv260617270C}, $T_{\rm dust}$ remains poorly constrained at $z\gtrsim6$, primarily due to the limited availability of sufficiently sensitive observations near the peak of the dust thermal emission.
Recent high-frequency ALMA observations start measuring $T_{\rm dust}$ directly for individual galaxies at $z\gtrsim6$ (see e.g., \citealt{2021MNRAS.508L..58B,2022MNRAS.515.1751W,2024MNRAS.527.6867A,2024MNRAS.533.3098A,2024ApJ...971..161M}). 

In addition to measuring $T_{\rm dust}$, understanding its relation to physical properties is essential, as dust heating is closely linked to the conditions of the ISM \citep{2007ApJ...657..810D,2022MNRAS.513.3122S}. 
While theoretical studies suggest that $T_{\rm dust}$ may correlate with metallicity or star formation surface density \citep[e.g.,][]{2019MNRAS.487.1844M,2019MNRAS.489.1397L,2020MNRAS.497..956S,2022MNRAS.513.3122S,2026arXiv260304505P}, such relationships have not yet been observationally demonstrated well, especially at high-$z$ \citep[see][as a local example]{2013A&A...557A..95R}.
In this regard, ALMA enables measurements of dust continuum emission across multiple wavelengths. 
We utilize ALMA Band 9 to observe dust continuum emission at a rest-frame wavelength of $\lambda_{\rm rest} \sim 63\,\mu{\rm m}$, in addition to measurements at $\sim 88\,\mu{\rm m}$ and $\sim 158\,\mu{\rm m}$. 
Observations at wavelengths closer to the peak of the dust thermal emission are crucial for constraining $T_{\rm dust}$.

Along with the dust continuum at $\lambda_{\rm rest} \sim 63\,\mu{\rm m}$, ALMA’s spectroscopic capability allows us to simultaneously observe the [O\,{\sc i}] 63\,$\mu$m line (hereafter \oi). 
Neutral oxygen (${\rm O}^0$) is one of the dominant coolants of neutral gas below an ionization potential of 13.6\,eV (comparable to that of hydrogen), and the \oi\ line traces dense gas owing to its high critical density \citep[$n_{\rm H, crit} \sim 10^5\,{\rm cm}^{-3}$,][]{1985ApJ...291..722T,1999ApJ...527..795K}. 
Therefore, \oi\ provides a unique probe of the dense neutral ISM in high-redshift galaxies \citep[e.g.,][]{2025arXiv250403831F}.
\oi\ lines are detected in local galaxies \citep{2015A&A...578A..53C,2017ApJ...846...32D,2018ApJ...861...95H}, but there are limited detections and constrains at $z\gtrsim4$ \citep{2020ApJ...889L..11R,2023RNAAS...7..188R,2025PASJ...77..139I}.

Several studies have reported elevated [O\,{\sc iii}]88$\mu$m/[C\,{\sc ii}]158$\mu$m ratio (hereafter \cii/\oiii) in galaxies at $z\gtrsim6$ (e.g., \citealp{2016Sci...352.1559I,2019PASJ...71...71H,2020ApJ...896...93H,2024ApJ...964..146F}, see also \citealp{2024MNRAS.527.6867A,2024MNRAS.532.2270B} for a sample selection effect), which are often interpreted as evidence for high ionization parameters, given their ionization potentials of 35.1\,eV and 11.3\,eV for \oiii\ and \cii, respectively (\citealp{2020ApJ...896...93H,2021MNRAS.505.5543V,2022ApJ...935..119S}, see also, \citealp{2022MNRAS.510.5603K,2025arXiv250712896P}). 
The critical densities of \oiii\ ($n_{\rm e, crit} \sim 510\,{\rm cm}^{-3}$) and \cii\ ($n_{\rm e, crit} \sim 45\,{\rm cm}^{-3}$ for electrons and $n_{\rm H, crit} \sim 2800\,{\rm cm}^{-3}$ for hydrogen atoms) are significantly lower than that of \oi. 
This difference enables us to investigate the role of gas density, in addition to ionization conditions, in driving the high \oiii/\cii\ ratios.

\renewcommand{\arraystretch}{1.1}
\tabcolsep = 0.07cm
%
%
%
%
%
%
\begin{deluxetable*}{ccccccccc}
\tablecaption{Summary of the sample \label{tab:tab1}}
\tablewidth{0pt}
\tablehead{\colhead{ID} & 
\colhead{$z_{\rm spec}$} & 
\colhead{$\mu$} &
\colhead{$M_{1500}$} &
\colhead{$12+\log{({\rm O/H})}$} &
\colhead{$\log M_{\ast}$} &
\colhead{$\log {\rm SFR}_{\rm IR+UV}$} &
\colhead{ALMA IDs} &
\colhead{ref}\\
 & & & ${\rm [mag]}$ & & [$M_{\odot}$] & [$M_{\odot}\,{\rm yr}^{-1}$] & &
}
\startdata
\multicolumn{8}{c}{{\bf main sample}} \\
J020038-021052 & 6.1120 & - & $\geq-21.5$ & - & - & $\geq2.58$ & \scriptsize{\#2022.1.00522.S, \#2023.1.00629.S} & 1\\
J091436+044231 & 5.8433 & - & $-23.57\pm0.04$ & - & - & $2.74_{-0.15}^{+0.27}$ & \scriptsize{\#2022.1.00522.S, \#2023.1.00629.S} & 1\\
J135348-001026 & 6.1702 & - & $-24.03\pm0.03$ & - & - & $2.96_{-0.09}^{+0.17}$ & \scriptsize{\#2022.1.00522.S, \#2023.1.00629.S} & 1\\
\hline
\multicolumn{8}{c}{{\bf supplemental sample ($T_{\rm dust}$, \oi)}} \\
J1211+0118 & 6.0293 & - & $-22.80\pm0.10$ & $8.51_{-0.15}^{+0.13}$ & $10.4\pm0.2$ & $1.80_{-0.08}^{+0.37}$ & \scriptsize{\#2017.1.00508.S, \#2023.1.01033.S, \#2023.1.00022.S} & 1,2,3\\
J0217+0208 & 6.2037 & - & $-23.12\pm0.05$ & $8.20_{-0.11}^{+0.15}$ & $10.2\pm0.2$ & $1.98_{-0.09}^{+0.25}$ & \scriptsize{\#2017.1.00508.S, \#2023.1.01033.S, \#2023.1.00022.S} & 1,2,3\\
\hline
\multicolumn{8}{c}{{\bf supplemental sample ($T_{\rm dust}$)}} \\
A1689-zD1 & 7.1332 & 9.6 & $-22.34\pm0.02$ & $8.36_{-0.10}^{+0.10}$ & $9.1\pm0.2$ & $1.72_{-0.05}^{+0.05}$ & 
\scriptsize{\makecell{\#2013.1.01064.S, \#2015.1.01406.S, \#2016.1.00954.S \\ \#2017.1.00775.S \#2019.1.01778.S}} & 5,6,7\\
B14-65666 & 7.1521 & - & $-22.29\pm0.18$ & $8.15_{-0.08}^{+0.07}$ & $9.8\pm0.2$ & $2.63_{-0.28}^{+0.91}$ & \scriptsize{\makecell{\#2015.1.00540.S, \#2016.1.00954.S, \#2017.1.00190.S \\ \#2018.1.01673.S, \#2019.1.01491.S \#2023.1.01033.S}} & 8,9,10\\
REBELS-25 & 7.3065 & - & $-21.53\pm0.05$ & $8.62_{-0.17}^{+0.17}$ & $9.1\pm0.1$ & $1.96_{-0.18}^{+0.72}$ & \scriptsize{\#2019.1.01634.L, \#2021.1.00318.S, \#2022.1.01324.S} & 11,12,13,14\\
MACS0416-Y1 & 8.3118 & 1.5 & $-21.15\pm0.02$ & $7.76_{-0.03}^{+0.03}$ & $9.0\pm0.1$ & $2.15_{-0.22}^{+0.61}$- & \scriptsize{\makecell{\#2016.1.00117.S, \#2017.1.00486.S, \#2017.1.00225.S \\ \#2019.1.00343.S, \#2024.1.00537.S}} & 15,16,17,18\\
\enddata
\vspace{3pt}
\vspace{-6pt}\tablenotetext{\small }{[References] 1) \citet{2024ApJ...971..161M}, 2) \citet{2020ApJ...896...93H}, 3) \citet{2025ApJ...993..204H}, 4) \citet{2019ApJ...883..183M}, 5) \citet{2015Natur.519..327W}, 6) \citet{2022ApJ...934...64A}, 7) \citet{2025arXiv251007936H}, 8) \citet{2019PASJ...71...71H}, 9) \citet{2024arXiv241215027J}, 10) \citet{2026arXiv260808015R}, 11) \citet{2022ApJ...931..160B}, 12) \citet{2024MNRAS.527.6867A}, 13)  \citet{2024MNRAS.533.3098A}, 14) \citet{2025arXiv250110559R}, 15) \citet{2015AandA...575A..92L}, 16) \citet{2019ApJ...874...27T}, 17) \citet{2024ApJ...977L..36H}, 18) \citet{2025MNRAS.544.1502B}, 19) \citet{2025AandA...696A..15R}
}
\vspace{-22pt}
\end{deluxetable*}

In this paper, we examine $T_{\rm dust}$ of the galaxies at $z\gtrsim6$ with ALMA band-9/10 observations and explore the connection between $T_{\rm dust}$ and the other properties.
Furthermore, we explore the physical conditions of the ISM, such as the ionization parameter ($U_{\rm ion}$) and gas density $n_{\rm H}$, through a combination of \oi, \oiii, and \cii\ lines, and investigate the origin of the high \oiii/\cii\ ratio at $z\gtrsim6$.
The paper is organized as follows: Section \ref{sec:sample} provides an overview of the datasets used in this work. 
Section \ref{sec:analysis} describes the method of measurements for dust continuum and emission line properties. 
In Section \ref{sec:results}, we report the results of $T_{\rm dust}$ and \oi\ line measurements and discuss key physical parameters determining $T_{\rm dust}$ and the origin of the high \Roiiicii\ ratio at $z\gtrsim6$. The conclusions are presented in Section \ref{sec:summary}. Throughout this paper, we assume a flat universe with the cosmological parameters of $\Omega_{\rm M}=0.3$, $\Omega_{\Lambda}=0.7$, $\sigma_{8}=0.8$, and $H_0=70$ \kms ${\rm Mpc}^{-1}$.

%
%
%
%
%

%
%
%
%
%
%
\section{Target, Observation and Data}\label{sec:sample}
\subsection{Main sample}\label{subsec:main}
The main targets of this paper are selected from the SERENADE (Systematic Exploration in the Reionization Epoch using Nebular And Dust Emission) survey, which is an ALMA program (ID:\#2022.1.00522.S, PI: Harikane) designed to observe the two brightest FIR fine structure lines (\cii\ and \oiii) in luminous LBGs at $z\sim6$.
The parent sample of the SERENADE survey is compiled from the literature, mostly from the galaxy sample identified in Hyper Suprime-Cam Subaru Strategic Program \citep[HSC-SSP;][]{2016ApJ...828...26M,2018PASJ...70S...4A,2018PASJ...70S..35M,2018ApJS..237....5M,2018PASJ...70S..10O,2022ApJS..259...20H}, with the aim of selecting galaxies that are likely normal star-forming systems and show no clear AGN or QSO signatures in their rest-frame UV spectra.
For details of the SERENADE survey design, we refer the reader to \citet{2024ApJ...971..161M} and Harikane et al. (in prep).

Among the 19 galaxies in the SERENADE sample, the main sample in this paper consists of three galaxies (J020038-021052, J091436-044231, and J135348-001026) that have been observed in an ALMA program \#2023.1.00629.S (PI: Mitsuhashi).
They are selected because of their potential high dust temperatures ($T_{\rm dust}\gtrsim60\,{\rm K}$), based on their rest-frame $158\,\mu{\rm m}$ and $88\,\mu{\rm m}$ observations \citep[][see Appendix \ref{appendix:sample_bias} for a potential sample bias]{2024ApJ...971..161M}.
The observations were conducted in ALMA Cycle 10 (September 2024) using Band 9.
The frequency setup was chosen to cover rest-frame $63\mu{\rm m}$ continuum and [O\,{\sc i}]63\,$\mu$m emission line (hereafter \oi).
Total on-source integration time depends on the source's brightness, ranging from 25--$80\,{\rm min}$.
The ALMA data were reduced using the standard pipeline within the Common Astronomy Software Application (CASA; \citealt{2022PASP..134k4501C}) versions adopted for the second-stage pipeline quality assurance (QA2).
All subsequent analyses were performed using CASA version 6.2.1.
The resulting beam sizes and RMS levels in the natural-weighted images range $0.28$--$0.35''$ and 4-10\,mJy\,${\rm beam}^{-1}$ in $10\,{\rm km}\,{\rm s}^{-1}$ channels, respectively.

%
%
%
%
%
%
\begin{figure*}[htbp]
\begin{center}
\epsscale{1.15}
\includegraphics[width=18cm,bb=0 0 1000 650, trim=0 1 0 0cm]{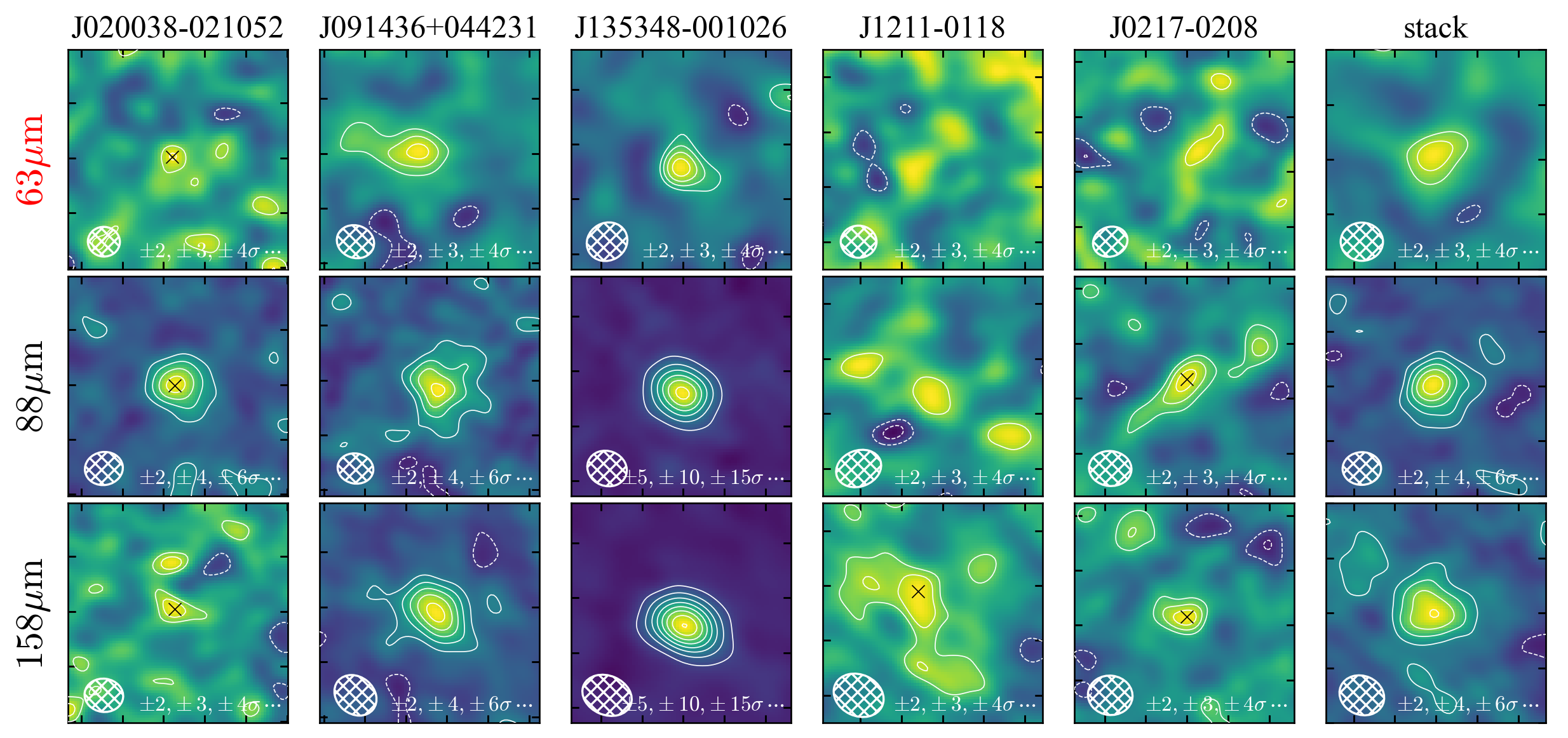}
\caption{Dust continuum maps of five galaxies and stacked image at the rest-frame $63\mu{\rm m}$ (top), $88\mu{\rm m}$ (middle), and $158\mu{\rm m}$ (bottom). The beam sizes and contour levels are shown in the bottom left and right, respectively. For sources with insufficient S/N for \texttt{imfit} in either band (${\rm S/N}<4.5$), the positions used for the flux measurements are marked with black crosses in the detected band when the dust continuum is detected (${\rm S/N}>3$).}
\label{fig:thumnail_dust}
\end{center}
\end{figure*}

\subsection{Supplemental sample}\label{subsec:supp}
We additionally incorporate two galaxies at $z\sim6$ with archival Band 9 observations covering the rest-frame $63\,\mu{\rm m}$ continuum and \oi\ (program ID:\#2023.1.01033.S, PI: Algera, hereafter referred to as the ($T_{\rm dust}$, \oi) supplemental sample). 
The two galaxies were originally reported in \citet{2020ApJ...896...93H}, and share similar properties with the main sample, such as absolute UV magnitude ($M_{\rm UV}$) and spectroscopic redshift \citep[see][]{2024ApJ...971..161M}.
These galaxies also have Band 10 observations covering the [O\,{\sc iii}]52\,$\mu${\rm m} emission line \citep[program ID: \#2023.1.00022.S, see][]{2025ApJ...993..204H}.

In addition, we incorporate four galaxies at $z=7.13$--8.31 with publicly available high-frequency ALMA observations covering rest-frame $\sim60\,\mu{\rm m}$ (hereafter referred to as the $T_{\rm dust}$) supplemental sample.
All of four galaxies have observations of \cii, \oiii\, and underlying continua \citep{2015Natur.519..327W,2017MNRAS.466..138K,2019PASJ...71...71H,2020MNRAS.495.1577I,2021ApJ...923....5S,2021MNRAS.508L..58B,2022MNRAS.515.1751W,2022ApJ...934...64A,2023MNRAS.518.6142A}.
The galaxies at $z=7.13$–8.31 are generally UV-bright systems similar to the $z\sim6$ galaxies and exhibit comparable UV and IR luminosities, although some of the $z>7$ sources are intrinsically faint and appear bright due to gravitational lensing (e.g., A1689-zD1).
A summary of the main and supplemental samples is provided in Table \ref{tab:tab1}.

For a fair comparison, we reanalyze the supplemental ALMA data in the same manner as applied to the main sample.
Any differences between our measurements and previous studies primarily arise from differences in the flux measurement methodology; however, our results are broadly consistent with those reported in the literature.

The supplemental sample has JWST observations covering rest-frame optical emission lines.
We adopt the gas-phase metallicity ($Z$) measurement of J1211-0118 and J0217-0208 \citep{2025ApJ...993..204H}, A1689-zD1 \citep{2025arXiv251007936H}, B14-65666 \citep{2024arXiv241215027J}, REBELS-25 \citep{2025arXiv250110559R}, and MACS0416-Y1 \citep{2024ApJ...977L..36H} based on the strong line calibrations.
The adopted metallicity values are summarized in Table \ref{tab:tab1}.

%
%
%
%
%
%
\section{Analysis}\label{sec:analysis}

%
%
%
%
%
%
\begin{figure*}[htbp]
\begin{center}
\epsscale{1.15}
\includegraphics[width=18cm,bb=0 0 1000 650, trim=0 1 0 0cm]{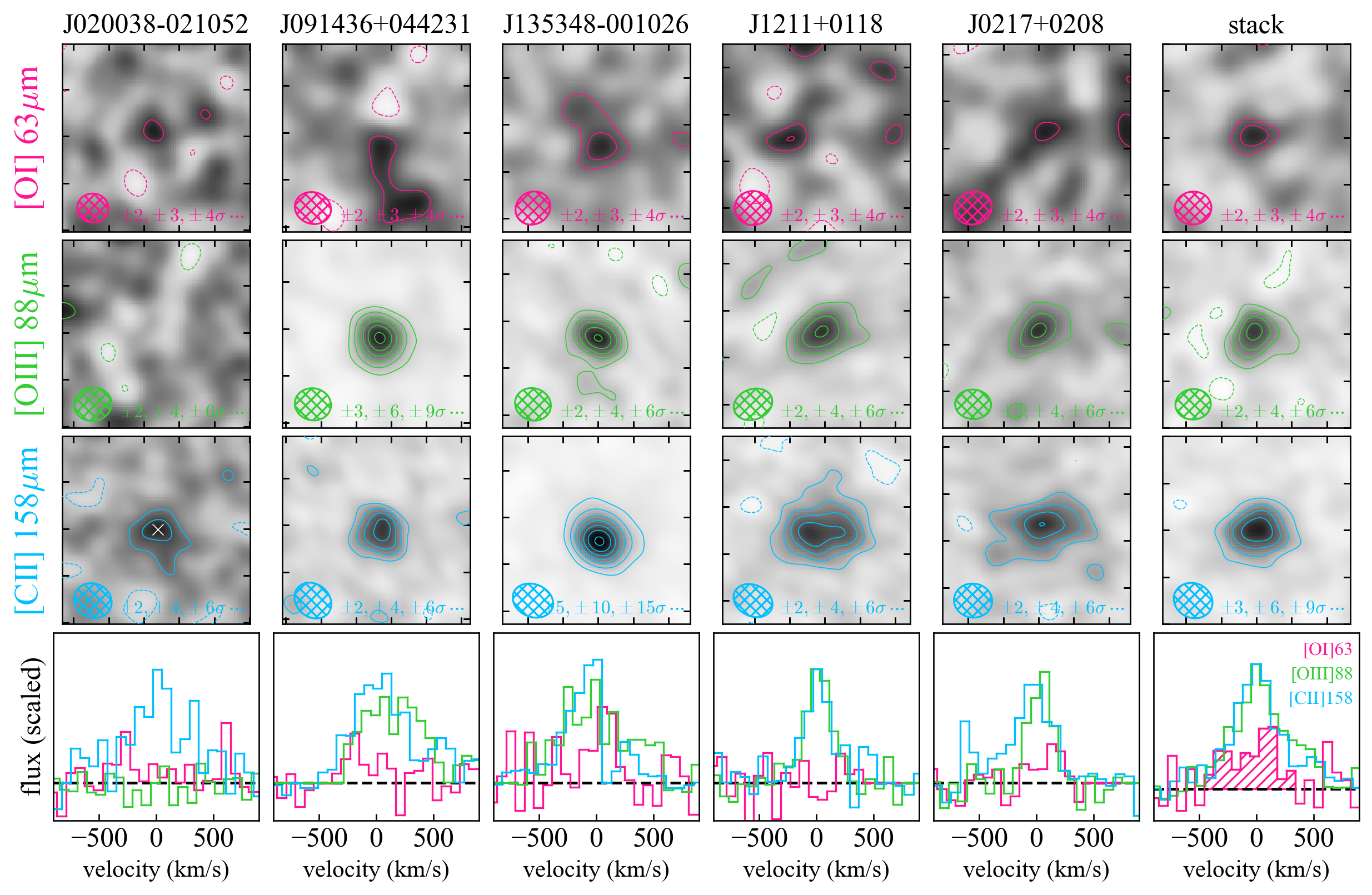}
\caption{[top three rows] Emission line maps of five main+supplemental galaxies and stacked image for \oi\ (1st row), \oiii\ (2nd row), and \cii\ (3rd row). The beam sizes and contour levels are shown in the bottom left and right, respectively. 
For sources with insufficient S/N for \texttt{imfit} in either band (${\rm S/N}<4.5$), the positions used for the flux measurements are marked with the white crosses in the detected line map when the line is detected (${\rm S/N}>3$).
[bottom row] The flux-scaled spectra of the \oi\ (red), \oiii\ (green), and \cii\ (blue).}
\label{fig:thumnail_line}
\end{center}
\end{figure*}

\subsection{Line and continuum fluxes}\label{subsec:flux}
In this section, we describe the detection and measurement of the dust continuum and emission line fluxes.
First, we make data cubes with natural weighting and apply a single Gaussian fit to the spectra extracted at the phase center to identify emission line features.
If any emission line feature is identified (i.e., the Gaussian fitting converges), we mask a frequency range $\pm2\times\text{full-width half maximum}$ (FWHM) from the central frequency using the CASA task \texttt{mstransform} to create dust-continuum visibility data.
For non-detections, we exclude frequencies within $\pm500\,{\rm km\,s^{-1}}$, which is much wider than the line width \citep[e.g.,][]{2020ApJ...896...93H}, from the expected central frequency at the galaxy's redshift to ensure eliminating potential emission line contamination to the dust continuum.
We also generate emission-line visibility data by selecting frequency ranges within the full width at tenth maximum (FWTM) to encompass the total line flux.
To place upper limits on the line fluxes, we also construct emission-line visibility data by assuming the same FWTM as that of \cii, since all galaxies in this paper have \cii\ detections and [O\,{\sc i}] lines have similar line widths to \cii\ \citep[e.g.,][]{2025PASJ...77..139I,2025arXiv250403831F}.

We then generate dust-continuum and emission-line maps using the CASA task \texttt{tclean}.
We reconstruct images with $uv$ tapers ranging from $0.1''$ to $1.0''$ in steps of $0.1''$, as well as naturally weighted images without tapering.
To evaluate the signal-to-noise ratio (S/N) of the target line or continuum, we obtain peak flux density within a $1.0''$ radius from each galaxy's central position.
Noise levels are estimated from images without primary-beam correction as the root-mean-square (RMS) of pixel values within the field of view where the primary-beam response exceeds 0.5.
If the highest S/N among the images in different taper scales exceeds ${\rm S/N}>3$, the dust continuum or emission line is considered detected.

Figure \ref{fig:thumnail_dust} presents the dust-continuum maps of the five galaxies in the main+supplemental ($T_{\rm dust}$, \oi) sample, as well as the stacked image (see Section \ref{subsec:stacking}).
All three galaxies in the main sample are detected in the rest-frame $158\mu{\rm m}$, $88\mu{\rm m}$, and $63\mu{\rm m}$.
One of the two galaxies in the supplemental ($T_{\rm dust}$, \oi) sample is detected at the rest-frame $63\,\mu{\rm m}$, while the other is not.
Both galaxies in the supplemental ($T_{\rm dust}$, \oi) sample are not detected in the rest-frame $52\,\mu{\rm m}$.
Among four galaxies in the supplemental ($T_{\rm dust}$) sample, three have detections at rest-frame $63\,\mu{\rm m}$ or $52\,\mu{\rm m}$ (see Figure \ref{fig:thumnail_dust_supp} in Appendix \ref{appendix:Tdfitting_supp}).
Figure \ref{fig:thumnail_line} shows the emission line maps and spectra. 
The \oi\ line is not clearly detected in individual images of the five galaxies in the main+ supplemental ($T_{\rm dust}$) samples, although J135348-001026 shows a tentative ($\sim4\sigma$) detection.

%
%
%
%
%
%
\begin{deluxetable*}{ccccccccc}
\tablecaption{Summary of the rest-frame FIR continuum properties \label{tab:tab2}}
\tablewidth{0pt}
\tablehead{\colhead{ID} & 
\colhead{$S_{158\mu{\rm m}}$} & 
\colhead{$S_{122\mu{\rm m}}$} &
\colhead{$S_{88\mu{\rm m}}$} & 
\colhead{$S_{63\mu{\rm m}}$} & 
\colhead{$S_{52\mu{\rm m}}$} &
\colhead{$T_{\rm dust}$} & 
\colhead{$\log L_{\rm IR}$\textsuperscript{\mbox{*}}} & 
\colhead{$\log M_{\rm dust}$\textsuperscript{\mbox{*}}}  \\
 & ${\rm [mJy]}$ & ${\rm [mJy]}$ & ${\rm [mJy]}$ & ${\rm [mJy]}$ & ${\rm [mJy]}$ & ${\rm [K]}$ & $[L_{\odot}]$ & $[M_{\odot}]$ 
}
\startdata
\multicolumn{9}{c}{{\bf main sample}} \\
J020038-021052 & $0.13\pm0.03$ & - & $0.51\pm0.06$ & $1.36\pm0.43$ & - & $85.6_{-24.7}^{+43.1}$ & $12.6_{-0.6}^{+0.5}$ & $6.4_{-0.2}^{+0.4}$ \\
J091436+044231 & $0.88\pm0.11$ & - & $4.35\pm0.59$ & $3.9\pm1.1$ & - & $49.9_{-11.9}^{+17.3}$ & $12.7_{-0.2}^{+0.2}$ & $8.0_{-0.3}^{+0.3}$ \\
J135348-001026 & $2.74\pm0.16$ & - & $7.97\pm0.23$ & $8.7\pm2.2$ & - &  $43.1_{-8.6}^{+12.2}$ & $12.9_{-0.1}^{+0.2}$ & $8.6_{-0.2}^{+0.3}$ \\
\hline
\multicolumn{9}{c}{{\bf supplemental sample ($T_{\rm dust}$, \oi)}} \\
J1211+0118 & $0.12\pm0.03$ & $0.18\pm0.02$ & $\leq0.50$ & $\leq1.43$ & $\leq1.24$ & $33.4_{-12.2}^{+29.6}$ & $11.3_{-0.3}^{+0.6}$ & $7.1_{-0.6}^{+0.6}$ \\
J0217+0208 & $0.13\pm0.03$ & $0.20\pm0.02$ & $0.31\pm0.07$ & $0.64\pm0.25$ & $\leq1.59$ & $47.2_{-14.8}^{+22.0}$ & $11.6_{-0.3}^{+0.4}$ & $7.1_{-0.5}^{+0.3}$ \\
\hline
\multicolumn{9}{c}{\bf supplemental sample ($T_{\rm dust}$)} \\
A1689-zD1$^{\ast}$ & $2.05\pm0.15$ & $1.09\pm0.02$ & $2.69\pm0.19$ & - & $1.84\pm0.43$ & $43.0_{-7.2}^{+9.1}$ & $11.4_{-0.1}^{+0.1}$ & $7.1_{-0.2}^{+0.2}$ \\
B14-65666 & $0.19\pm0.03$ & $0.22\pm0.01$ & $0.65\pm0.14$ & $1.80\pm0.56$ & - & $80.5_{-20.6}^{+56.6}$ & $12.6_{-0.5}^{+0.5}$ & $6.6_{-0.2}^{+0.3}$ \\
REBELS-25 & $0.18\pm0.02$ & - & $0.60\pm0.14$ & - & $\leq1.71$ & $43.3_{-16.2}^{+24.3}$ & $11.9_{-0.3}^{+0.5}$ & $7.4_{-0.5}^{+0.4}$ \\
MACS0416-Y1$^{\ast}$ & $\leq0.033$ & - & $0.092\pm0.009$ & $0.226\pm0.064$ & $0.46\pm0.16^{\S}$ & $100.7_{-12.1}^{+43.7}$ & $12.1_{-0.3}^{+0.4}$ & $5.8_{-0.3}^{+0.1}$ \\
\hline
\multicolumn{9}{c}{{\bf stack}} \\
five galaxies$^{\dagger}$ & $0.60\pm0.04$ & - & $1.23\pm0.08$ & $1.99\pm0.64$ & - & $39.8_{-7.6}^{+10.6}$ & $11.8_{-0.1}^{+0.1}$ & $7.7_{-0.3}^{+0.2}$ \\
four galaxies$^{\ddagger}$ & $0.13\pm0.02$ & - & $0.45\pm0.04$ & $0.71\pm0.18$ & - & $54.7_{-13.8}^{+29.7}$ & $11.8_{-0.2}^{+0.4}$ & $7.0_{-0.4}^{+0.3}$ \\
\hline
\enddata
\tablecomments{
\vspace{-6pt}\tablenotetext{\small \S}{Rest-frame $\sim45\,\mu{\rm m}$ flux density, see \citet{2025MNRAS.544.1502B}}
\vspace{-6pt}\tablenotetext{\small \ast}{$M_{\rm dust}$ and $L_{\rm IR}$ values are corrected for the gravitational magnification factor}
\vspace{-6pt}\tablenotetext{\small \dagger}{Stacking of the 5 galaxies in main sample and supplemental sample ($T_{\rm dust}$,\oi)}
\vspace{-6pt}\tablenotetext{\small \ddagger}{Stacking of the 4 galaxies excluding J135348-001026, based on its tentative \oi\ detection}
}
\end{deluxetable*}
%
%
%
%
%

To ensure consistent flux measurements across different ALMA bands, we derive line and continuum fluxes using the following two methods: (1) For the galaxies detected in all continuum or emission line maps, we perform two-dimensional Gaussian fitting using {\sc CASA/imfit}, (2) For sources not detected in either bands or lines, we select maps with similar synthesized beam sizes (within $\pm10\%$) and measure the peak flux density and RMS noise to uniformly evaluate the fluxes and noises contained within a single synthesized beam.
The typical resulting synthesized beam used for (2) is $\sim0.6$--$0.8''$, which is sufficiently larger than the sizes of the galaxies ($\sim0.2$--$0.3''$).
To avoid potential flux misestimation in {\sc imfit} due to mismatches between the dirty and clean beams, we apply modest $uv$ tapers for some sources prior to running {\sc imfit} (e.g., A1689-zD1).
The measured dust continuum and emission line fluxes are listed in Tables \ref{tab:tab2} and \ref{tab:tab3}.
Although it is not included in the values in the tables, 10\% (or 20\% for Band-9 and 10) systematic flux calibration uncertainties are added in quadrature to the measured uncertainties in the following calculations.

\tabcolsep = 0.1cm
%
%
%
%
%
%
\begin{deluxetable*}{cccccccccc}
\tablecaption{Summary of the rest-frame FIR line properties \label{tab:tab3}}
\tablewidth{0pt}
\tablehead{\colhead{ID} & 
\colhead{${\rm FWHM}_{\text{\cii}}$} & 
\colhead{$S\Delta v_{\text{\cii}}$} &
\colhead{$L_{\text{\cii}}$} & 
\colhead{${\rm FWHM}_{\text{\oiii}}^{\S}$} & 
\colhead{$S\Delta v_{\text{\oiii}}$} &
\colhead{$L_{\text{\oiii}}$} &  
\colhead{${\rm FWHM}_{\text{\oi}}^{\S}$} & 
\colhead{$S\Delta v_{\text{\oi}}$} &
\colhead{$L_{\text{\oi}}$} \\
 & $[{\rm km}\,{\rm s}^{-1}]$ & $[{\rm Jy.}\,{\rm km}\,{\rm s}^{-1}]$ & $[\times10^9 L_{\odot}]$ & $[{\rm km}\,{\rm s}^{-1}]$ & $[{\rm Jy.}\,{\rm km}\,{\rm s}^{-1}]$ & $[\times10^9 L_{\odot}]$ & $[{\rm km}\,{\rm s}^{-1}]$ & $[{\rm Jy.}\,{\rm km}\,{\rm s}^{-1}]$ & $[\times10^9 L_{\odot}]$
}
\startdata
\multicolumn{10}{c}{{\bf main sample}} \\
J020038-021052 & $647\pm80$ & $0.76\pm0.18$ & $0.74\pm0.18$ & - & $\leq1.03$ & $\leq1.79$ & - & $\leq1.03$ & $\leq9.64$ \\
J091436+044231 & $510\pm41$ & $1.71\pm0.27$ & $1.55\pm0.25$ & $828\pm170$ & $8.07\pm0.52$ & $13.0\pm0.8$ & - & $\leq1.55$ & $\leq6.13$ \\
J135348-001026 & $486\pm12$ & $4.41\pm0.17$ & $4.32\pm0.17$ & $815\pm58$ & $8.85\pm1.04$ & $15.5\pm1.8$ & $237\pm59$ & $9.84\pm2.42$ & $24.1\pm5.9$ \\
\hline
\multicolumn{10}{c}{{\bf supplemental sample ($T_{\rm dust}$, \oi)}} \\
J1211+0118 & $186\pm15$ & $0.64\pm0.08$ & $0.61\pm0.07$ & $269\pm31$ & $2.24\pm0.34$ & $3.79\pm0.57$ & - & $\leq1.01$ & $\leq2.08$ \\
J0217+0208 & $434\pm37$ & $0.73\pm0.11$ & $0.72\pm0.11$ & $230\pm17$ & $1.57\pm0.28$ & $2.76\pm0.49$ & - & $\leq0.84$ & $\leq3.56$ \\
\hline
\multicolumn{10}{c}{\bf supplemental sample ($T_{\rm dust}$)} \\
A1689-zD1$^{\ast}$ & $309\pm4$ & $1.65\pm0.08$ & $1.99\pm0.09$ & $353\pm15$ & $1.16\pm0.04$ & $2.5\pm0.1$ & - & - & - \\
B14-65666 & $269\pm15$ & $0.92\pm0.10$ & $1.12\pm0.12$ & $464\pm44$ & $2.30\pm0.21$ & $4.99\pm0.45$ & - & - & - \\
REBELS-25 & $375\pm3$ & $1.67\pm0.03$ & $2.08\pm0.04$ & $289\pm42$ & $1.93\pm0.54$ & $4.31\pm1.22$ & - & - & - \\
MACS0416-Y1$^{\ast}$ & $181\pm26$ & $0.16\pm0.03$ & $0.23\pm0.05$ & $186\pm14$ & $0.72\pm0.06$ & $1.94\pm0.15$ & - & - & - \\
\hline
\multicolumn{10}{c}{{\bf stack }} \\
five galaxies$^{\dagger}$ & $600\pm41$ & $1.24\pm0.07$ & $1.18\pm0.07$ & $493\pm38$ & $1.97\pm0.23$ & $3.35\pm0.40$ & $544\pm153$ & $1.38\pm0.34$ & $3.32\pm0.82$ \\
four galaxies$^{\ddagger}$ & $421\pm36$ & $0.89\pm0.08$ & $0.85\pm0.07$ & $527\pm46$ & $1.80\pm0.23$ & $3.07\pm0.39$ & $564\pm165$ & $1.23\pm0.33$ & $2.95\pm0.80$ \\
\hline
\enddata
\tablecomments{
\vspace{-6pt}\tablenotetext{\small \S}{${\rm FWHM}_{\rm [OI]63}$ and ${\rm FWHM}_{\rm [OIII]88}$ are assumed to be same with ${\rm FWHM}_{\rm [CII]158}$ in case of non-detection}
\vspace{-6pt}\tablenotetext{\small \ast}{The values are corrected for the gravitational magnification factor ($\mu=9.6$ for A1689-zD1 and $\mu=1.5$ for MACS0416-Y1)}
\vspace{-6pt}\tablenotetext{\small \dagger}{Stacking of the 5 galaxies in main sample and supplemental sample ($T_{\rm dust}$,\oi)}
\vspace{-6pt}\tablenotetext{\small \ddagger}{Stacking of the 4 galaxies excluding J135348-001026, owing to its tentative \oi\ detection}
}
\end{deluxetable*}
%
%
%
%
%

\subsection{Stacking analysis}\label{subsec:stacking}

In this section, we describe the stacking analysis of the dust continuum and emission lines to obtain average properties.
We perform stacking for five galaxies at $z\sim6$ (J020038-021052, J091436+044231, J135348-001026, J1211+0118, and J0217+0208) that have observations of \oi, \oiii, \cii, and underlying continua.
We also test an alternative stacking excluding J135348-001026, for which \oi\ is tentatively detected and may bias the average \oi\ luminosity.
Indeed, J135348-001026 shows significantly brighter fluxes than the other four galaxies included in the stacking analysis.
Hereafter, we mainly use this stacking result without J135348-001026, but note that there is no major impact on the conclusion if we refer to the results with all five galaxies (within $\sim20\%$, see Tables \ref{tab:tab2} and \ref{tab:tab3}).
To account for source-to-source variations in beam size, we perform the stacking in the visibility domain.
We use the CASA task \texttt{fixvis} to align source coordinates to (00h00m00.00s, 00d00m00.0s) and combine the data using the CASA task \texttt{concat}.
Imaging and flux measurements follow the procedure described in Section \ref{subsec:flux}.

The stacked dust-continuum and emission-line maps are shown in the rightmost panels in Figure \ref{fig:thumnail_dust} and \ref{fig:thumnail_line},
and the stacked spectra are also shown in Figure \ref{fig:thumnail_line} for visualization purposes.
The stacked continuum and emission-line maps show clear detections at the $>4\sigma$ level at all three bands ($63\,\mu{\rm m}$, $88\,\mu{\rm m}$, $158\,\mu{\rm m}$) and lines (\oi, \oiii, and \cii), coincident with the positions of the emission seen at other wavelengths or in other lines.
The measured stacked line and continuum fluxes are also listed in Tables \ref{tab:tab2} and \ref{tab:tab3}.

%
%
%
%
%
%
\begin{figure*}[!htbp]
\begin{center}
\centering
\includegraphics[width=0.9\textwidth]{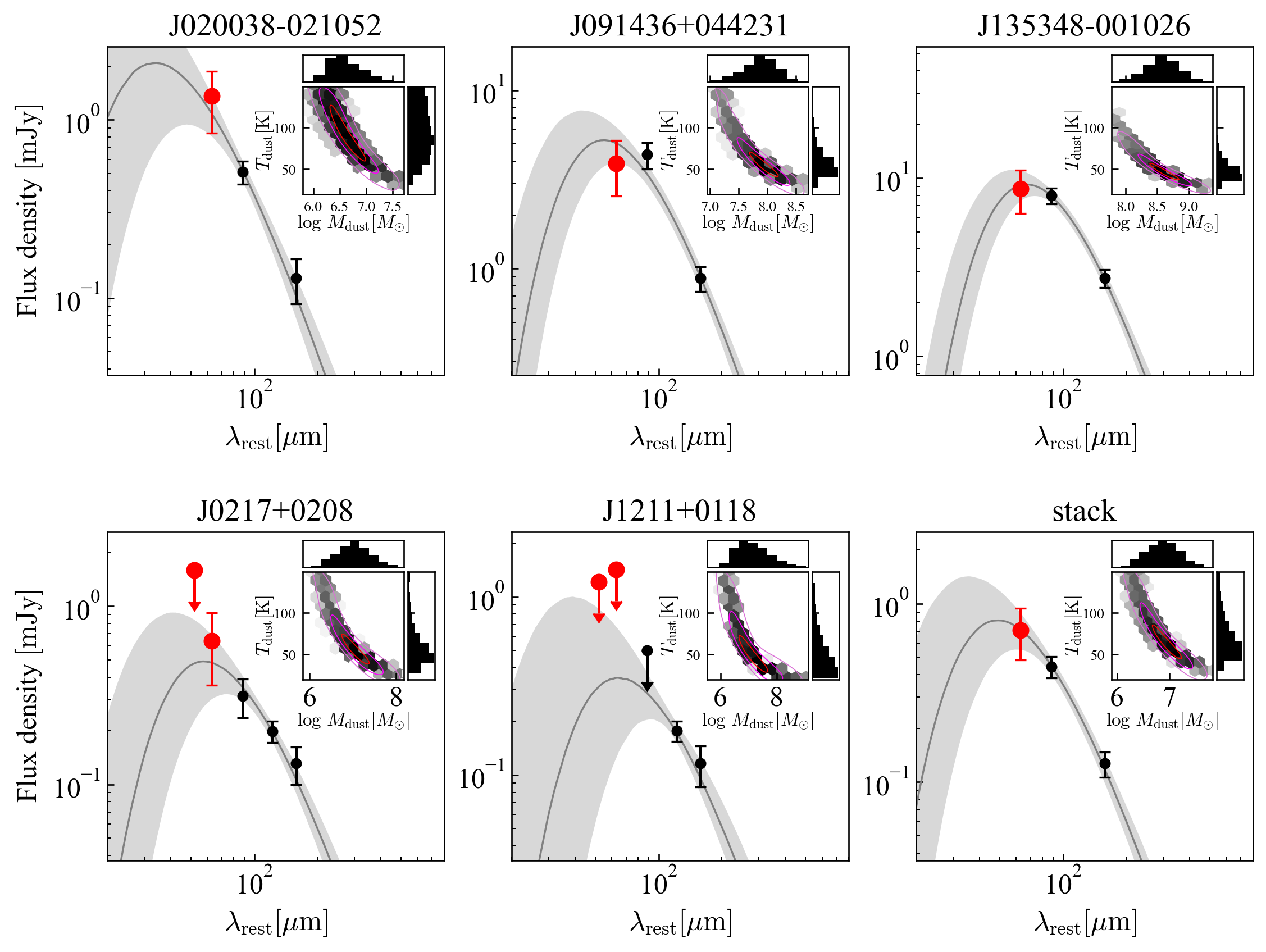}
\caption{Optically thin MBB fitting results of five individual galaxies and stacked average. New constraints from Band-9 and 10 observations are shown in the red circles. Each panel shows the 1$\sigma$ confidence interval for the MBB profiles as a function of the rest-frame wavelength (shaded area). The observed fluxes and 3$\sigma$ upper limits are shown in the black circles. The inset panels illustrate the posterior distributions of the MCMC procedure with 1-,2-, and 3-$\sigma$ contours. The best-fit and 1$\sigma$ uncertainties of $L_{\rm IR}$ are shown inside the left panels.}
\label{fig:Tdfitting}
\end{center}
\end{figure*}

\subsection{$T_{\rm dust}$ and $M_{\rm dust}$ estimation}\label{subsec:mcmc}
We fit a modified blackbody (MBB) profile to constrain the properties of the dust emission.
We basically follow the methodology used in \citet{2024ApJ...971..161M}, and summarize it briefly here.

The MBB profile is primarily characterized by three parameters: dust temperature ($T_{\rm dust}$), dust mass ($M_{\rm dust}$), and emissivity of the dust grain ($\beta_{\rm dust}$).
Under the optically thin assumption, the observed MBB flux density at $\nu_{\rm obs}$, taking into account the cosmic microwave background (CMB) effect based on \citet{2013ApJ...766...13D}, can be represented as follows:

\begin{equation}\label{eq1}
F_{\nu_{\rm obs}}\!=\! \left(\frac{1+z}{4\pi d_{\rm L}^2}\right)M_d\kappa_{\nu}[B_{\nu}(T_{\rm dust})-B_{\nu}(T_{{\rm CMB},z})].
\end{equation}

\noindent Here $d_{\rm L}$ denotes the luminosity distance at redshift $z$.
$B_{\nu}(T_{\rm dust})$ and $B_{\nu}(T_{\rm CMB})$ represents the blackbody radiation at the temperature of $T_{\rm dust}$ and $T_{{\rm CMB},z}$ ($\equiv T_{{\rm CMB},z=0}\times[1+z]$), respectively.
The absorption coefficient $\kappa_{\nu}$ is parameterized as  $\kappa_{\nu}=\kappa_{\ast}(\nu/\nu_{\ast})^{\beta_{\rm dust}}$ with the normalization of Milky Way value, [$\kappa_{\ast}$, $\nu_{\ast}$]=[10.41 cm$^{2}$g$^{-1}$, 1900 GHz] \citep[e.g.,][]{2021MNRAS.508L..58B,2023MNRAS.518.6142A,2022MNRAS.512...58F}.

In this work, we use the optically-thin assumption (eq \ref{eq1}) to estimate $T_{\rm dust}$, $\beta_{\rm dust}$, $M_{\rm dust}$ in the same manner as previous studies \citep[e.g.,][]{2020MNRAS.494.3828D,2021ApJ...919...30D,2021MNRAS.508L..58B,2023MNRAS.518.6142A,2024ApJ...971..161M}.
We note that the resulting infrared luminosity ($L_{\rm IR}$) is largely insensitive to the assumption of optical thickness, whereas the inferred $T_{\rm dust}$ becomes higher when adopting an optically thick MBB model with $\lambda_0=100\,\mu$m \citep[see also,][]{2022MNRAS.512...58F}.

We employ a Markov Chain Monte Carlo (MCMC) approach, using the \texttt{emcee} library, to fit the MBB models to the measured dust-continuum flux densities.
We adopt a logarithmically uniform prior on the dust masses with a range of $\log M_{\rm dust}\,[M_{\odot}]\in[4,10]$ and linearly uniform prior on the dust temperatures with $T_{\rm dust}\,[{\rm K}]\in[T_{{\rm CMB},z},150]$, respectively.
Since the sampling range of the FIR SED is not enough to constrain $\beta_{\rm dust}$, we adopt a Gaussian prior with a mean value of $\langle\beta_{\rm dust}\rangle=1.8$ and a standard deviation of $\sigma_{\beta_{\rm dust}}=0.5$ \citep{2014A&A...571A..11P,2014MNRAS.440..942C,2019MNRAS.489.4389L,2018MNRAS.475.2097C,2021ApJ...919...30D,2023MNRAS.523.3119W}.
The upper limits are treated as in \citet{2012PASP..124.1208S}, where they introduce the penalizing term depending on the model flux based on the Gaussian probability distribution.
We compute best-fit values and 1$\sigma$ uncertainties from the modes with the highest posterior density intervals.

Figure \ref{fig:Tdfitting} shows the results of the MBB fittings for the five main+supplemental ($T_{\rm dust}$,\oi) targets as well as the stacked averages.
The infrared luminosities are computed by integrating the MBB profile over $8$--$1000\,\mu{\rm m}$.
Our MBB fitting successfully constrains $T_{\rm dust}$, $M_{\rm dust}$, and the resulting $L_{\rm IR}$.
The derived infrared luminosities of the three main targets span $\log L_{\rm IR}\,[L_{\odot}]=12.6$--12.9, comparable to those of local ULIRGs.
The fitting results are summarized in Table \ref{tab:tab2}.
The fitting results for the supplemental ($T_{\rm dust}$) sample are shown in Appendix \ref{appendix:Tdfitting_supp}.

%
%
%
%
%
%
\begin{figure}[t]
\begin{center}
\epsscale{1.15}
\includegraphics[width=8.5cm,bb=0 0 200 150, trim=0 1 0 0cm]{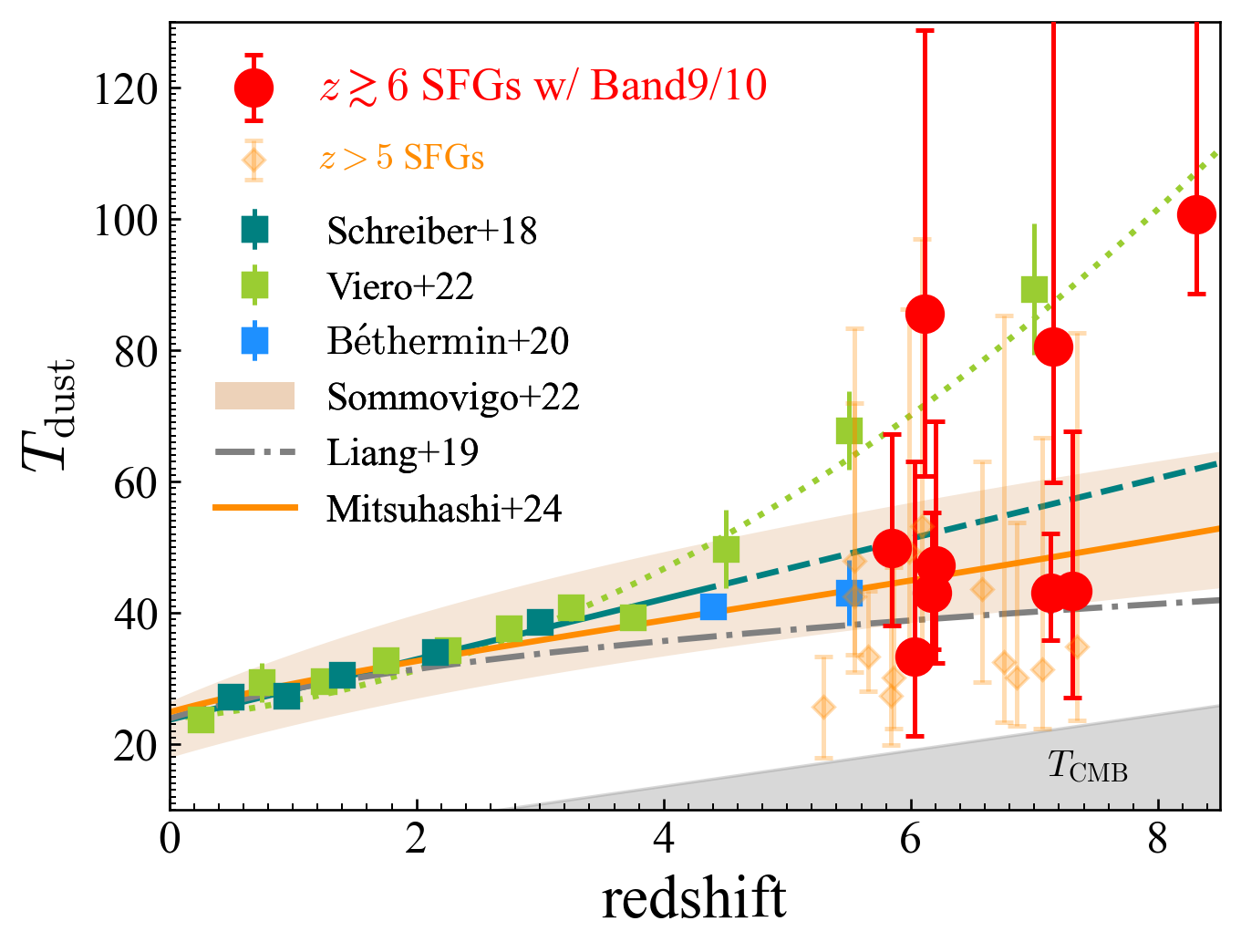}
\caption{Dust temperature as a function of redshift. The galaxies with Band-9/10 observations analysed in this work are shown as red circles \citep[see also,][]{2025MNRAS.544.1502B}.
The other measurements for $z\gtrsim5$ galaxies in \citet{2024ApJ...971..161M} are shown as orange markers \citep[see also,][]{2019PASJ...71...71H,2019MNRAS.487L..81L,2020MNRAS.498.4192F,2021ApJ...923....5S,2020MNRAS.493.4294B,2021MNRAS.508L..58B,2022MNRAS.515.1751W,2024MNRAS.527.6867A}. 
For comparison, previous results based on stacking analysis and the expected redshift evolution are shown in blue, light blue, and green colors \citep{2018A&A...609A..30S,2020A&A...643A...2B,2022MNRAS.516L..30V}.
The model predictions from \citet{2019MNRAS.489.1397L}, \citet{2022MNRAS.513.3122S}, and  \citet{2024ApJ...971..161M} are shown in gray line, brown shaded region, and orange line, respectively.
}
\label{fig:zTdust}
\end{center}
\end{figure}
%
%
%
%
%
%

%
%
%
%
%
%
\section{Results and Discussion}\label{sec:results}
\subsection{High $T_{\rm dust}$ galaxies in the early Universe}

In Figure \ref{fig:zTdust}, we show $T_{\rm dust}$ as a function of redshift. 
Among nine galaxies with rest-frame coverage down to $\lesssim63\,\mu{\rm m}$ at $z\gtrsim6$, six have $T_{\rm dust}\sim40$--50\,K, consistent with the expected redshift evolution reported in previous studies \citep[][]{2018A&A...609A..30S,2019MNRAS.489.1397L,2022MNRAS.513.3122S,2024ApJ...971..161M}.
The remaining three galaxies exhibit higher dust temperatures ($T_{\rm dust}\gtrsim70\,{\rm K}$, see the posterior distribution in the inset panel of Figure \ref{fig:Tdfitting}), as indicated by their high $S_{63\mu{\rm m}}/S_{88\mu{\rm m}}$ ratios.
The posterior distributions of $T_{\rm dust}$ for these galaxies almost completely rule out $T_{\rm dust}<50\,{\rm K}$.
Constraints from rest-frame $63\,\mu{\rm m}$ measurements are critical for precisely determining $T_{\rm dust}$, in particular for distinguishing between $T_{\rm dust}\lesssim60\,{\rm K}$ and $\gtrsim60\,{\rm K}$.
In the following, we examine the possible origin of the $T_{\rm dust}$ variation at the fixed redshift and the $T_{\rm dust}$ evolution across the $z\sim0$ to $z\sim6$--9.

\subsubsection{$T_{\rm dust}$ variation}\label{subsec:tdustvar}

%
%
%
%
%
%
\begin{figure*}[htbp]
\begin{center}
\epsscale{1.15}
\includegraphics[width=18cm,bb=0 0 1000 650, trim=0 1 0 0cm]{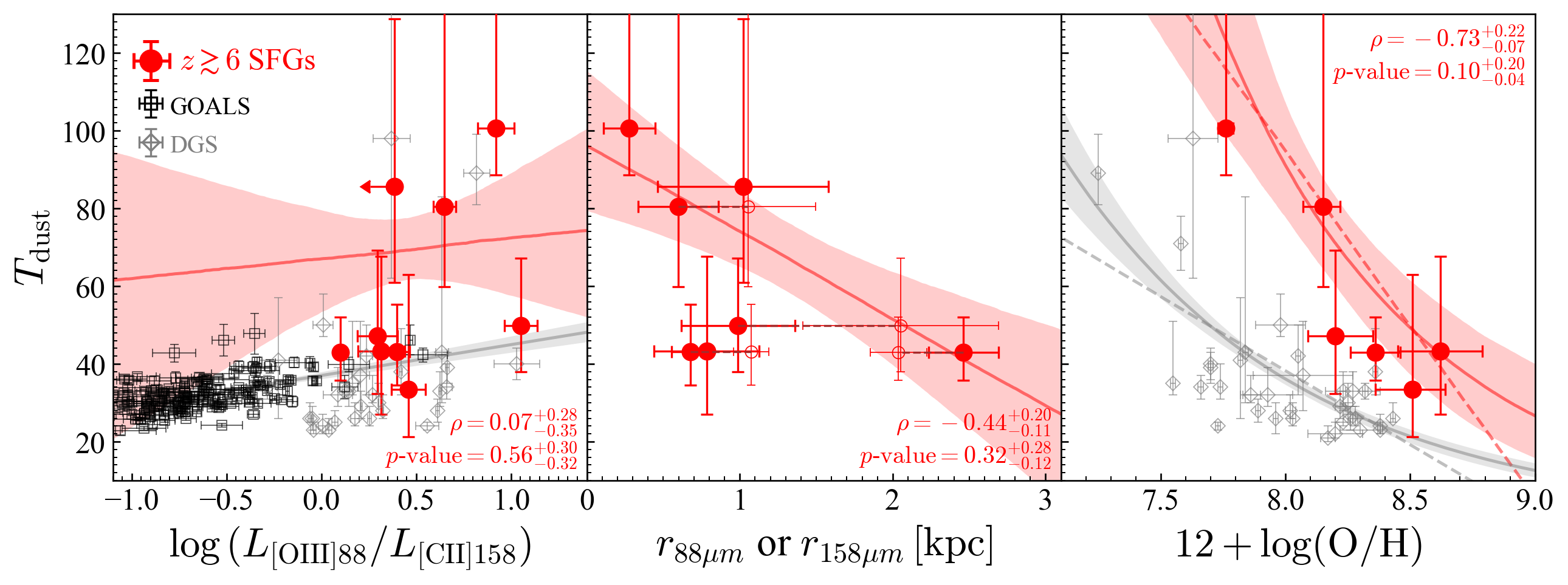}
\caption{Dust temperature as a function of $\log (L_{\rm [OIII]88}/L_{\rm [CII]158)}$ (left), dust continuum size at $\lambda_{\rm rest}=88\,\mu{\rm m}$ ($r_{88\mu{\rm m}}$) or $158\,\mu{\rm m}$ ($r_{158\mu{\rm m}}$, middle), and metallicity (right).
The $z\gtrsim6$ galaxies analysed in this work are shown in the red markers, and local samples from DGS \citep{2015A&A...578A..53C} and GOALS \citep{2013ApJ...774...68D} are shown in the gray diamonds and black squares as a comparison.
Pearson's correlation and $p$-values are included in each panel with their uncertainties derived from bootstrapping.
The linear fitting for $z\gtrsim6$ and the local samples are illustrated in red and gray curves.
In the right panel, we show a linear form ($T_{\rm dust}\propto \log Z$, dashed lines) and the log form ($\log T_{\rm dust}\propto \log Z$, solid lines).
In the middle panel, we show the size of the dust continuum ($r_{88\mu{\rm m}}$ or $r_{158\mu{\rm m}}$) with a higher S/N in the filled markers, and the others in the open markers.
The Pearson's correlation and fitting results do not change if we use either $r_{88\mu{\rm m}}$ or $r_{158\mu{\rm m}}$.
Among three comparisons, only $Z$ exhibits tentative correlation with $T_{\rm dust}$.}
\label{fig:Tdcorrelation}
\end{center}
\end{figure*}

Figure \ref{fig:zTdust} demonstrates substantial galaxy-to-galaxy variability in $T_{\rm dust}$.
To explore possible drivers of the $T_{\rm dust}$ variations, we test correlations between $T_{\rm dust}$ and several observables using a Pearson correlation analysis (Figure \ref{fig:Tdcorrelation}).

(1) \oiii-to-\cii\ luminosity ratio, \Roiiicii.
Previous studies have suggested that galaxies with low \Roiiicii\ tend to exhibit lower dust temperatures \citep[e.g.,][]{2018ApJ...869L..22W,2024MNRAS.527.6867A}.
We find no statistically significant correlation between $T_{\rm dust}$ and \logRoiiicii\ within the current dynamic range and sample size in the $z\gtrsim6$ sample, while the local sample shows a positive correlation.
As shown in \citet{2020ApJ...896...93H}, high-$z$ galaxies tend to exhibit \Roiiicii\ values more than twice those of local samples, and therefore the observable dynamic range of \Roiiicii\ is small.
Additionally, \Roiiicii\ is known to correlate with several ISM parameters, such as metallicity, gas density, ionization parameter, and PDR covering fraction (see Section \ref{subsec:OICIIOIII} for more discussion about the \Roiiicii).
This limited dynamic range, combined with the complex physical dependencies of \Roiiicii, is likely to make the correlation unclear.

(2) dust continuum size, $r_{\rm dust}$.
Dust continuum sizes are expected to be small in compact starburst systems, which are often associated with higher $T_{\rm dust}$ \citep[e.g.,][]{2007A&A...462...81C,2013ApJ...774...68D}.
If the compact starburst activity drives high $T_{\rm dust}$, there may be a correlation between $T_{\rm dust}$ and the size ratio.
Here, we use \texttt{uvmultifit} to measure sizes.
We fit the elliptical profile (or circular profile if ellipticity is not constrained well) to the visibility data produced in Section \ref{subsec:flux} for galaxies detected with sufficient S/N ($>4.5\sigma$) to allow reliable size measurements.
We use the circularized radii as a representation of the dust continuum sizes.
When dust continuum emission is detected at both rest-frame $88\,\mu{\rm m}$ and $158\,\mu{\rm m}$, we adopt the measurement at the wavelength with higher S/N.
We note that the fluxes derived from \texttt{uvmultifit} and our measurements in Section \ref{subsec:flux} are consistent (see Appendix in \citealt{2024ApJ...971..161M}).

Although some high $T_{\rm dust}$ sources show potential compact dust size, we do not find any significant correlation between $T_{\rm dust}$ and the size.
Therefore, compact starburst activity is not likely to be the main driver of the $T_{\rm dust}$ variations.
It is worth noting that simple characterization of the single disk component may prevent proper understanding of the correlation between $T_{\rm dust}$ and size.
For instance, sizes in the multi-component system may be overestimated in a simple disk modeling, while each component is compact.
Indeed, one extended object in the middle panel of Figure \ref{fig:Tdcorrelation} is A1689-zD1, which is known as a multi-component, merging system \citep{2025arXiv251007936H,2025A&A...701A..85K}.

(3) gas-phase metallicity, $Z$.
Analytical models predict that gas-phase metallicity is one of the key factors determining $T_{\rm dust}$ \citep{2022MNRAS.513.3122S}.
We find a tentative negative correlation with a $p$-value of $\sim0.09$ with $N=6$.
If this tentative correlation is real, the inferred power-law slope of the $T_{\rm dust}$–$Z$ relation in the $z\gtrsim6$ sample ($-0.54\pm0.22$) is consistent with that derived for the local DGS sample ($-0.46\pm0.07$, \citealp[see also,][]{2013A&A...557A..95R}) within $1\sigma$ uncertainties.
The inferred slope is steeper than the $Z^{-1/(4+\beta_{\rm dust})}$ dependence predicted by \citet{2022MNRAS.513.3122S} at $\sim2\sigma$ level in a reasonable $\beta_{\rm dust}$ range for high-$z$ galxies \citep[$\beta_{\rm dust}\sim1.8-2.0$,][]{2020RSOS....700556H,2023MNRAS.523.3119W}.
In \citet{2022MNRAS.513.3122S}, the dependence on the $Z$ comes from the dust-to-gas mass ratio (D/G) proportional to $Z$ \citep[see also,][]{2026arXiv260304505P}.
Even taking the stronger $Z$ dependence on D/G in low-$Z$ environments \citep[$Z\lesssim0.2\,Z_{\odot}$,][]{2014A&A...563A..31R} into account, the increase in $T_{\rm dust}$ is approximately a factor of $\sim1.3$--1.5, which is insufficient to explain the observed trend.

The potential strong dependence on $Z$ may reflect inefficient dust shielding and harder stellar SED in low-metal environments. 
Reduced dust shielding in low-metallicity environments allows far-UV photons to penetrate deeper into the ISM, exposing a larger fraction of the dust mass to elevated radiation fields \citep{1999ApJ...513..275B}.
Stars formed in low-metal environments have high effective temperatures \citep[e.g.,][]{2016MNRAS.456..485S}.
Both effects lead to stronger FUV irradiation of dust surrounding star-forming regions, thereby enhancing the average $T_{\rm dust}$ across the galaxy.
Such additional factors may accelerate the dependence of $T_{\rm dust}$ on $Z$.

We note that accurately measuring $Z$ is also crucial.
\citet{2025ApJ...993..204H} introduced a 2-zone ISM structure to explain \oiii/[O\,{\sc iii}]$\lambda5007$ \citep[see also,][]{2025ApJ...991L..38U}, and found the metallicity depends on the assumption of a 1-zone or 2-zone ISM.
The metallicities adopted in this work (Table \ref{tab:tab1}) are derived from optical strong-line diagnostics ([O\,{\sc iii}]$\lambda5007/{\rm H}\beta$ and [O\,{\sc iii}]$\lambda5007$/[O\,{\sc ii}]$\lambda\lambda3727,3729$), which primarily probe relatively dense ionized gas. 
In contrast, FIR [O\,{\sc iii}] emission may arise from more diffuse ionized gas that occupies a larger fraction of the ISM volume. 
If the metallicity of the diffuse ISM differs from that of the dense ISM by up to $\sim0.5,{\rm dex}$, as suggested by \citet{2025ApJ...993..204H}, the metallicity inferred from optical emission lines may not accurately represent the metallicity of the dust-emitting gas. 
Such a mismatch could weaken any intrinsic correlation between $T_{\rm dust}$ and $Z$, although it remains unclear which gas phase is most closely associated with the dust properties.

Interestingly, one high $T_{\rm dust}$ galaxy ($T_{\rm dust}\sim90\,{\rm K}$), J020038-021052, shows bright Ly$\alpha$ emission ($EW_{\rm Ly\alpha}\sim500\,$\AA, Ono et al. in prep), suggesting a metal-poor condition or significant AGN contribution \citep{2003A&A...397..527S,2017MNRAS.465.1543H}.
MACS0416-Y1 also shows an indication of possible AGN activity \citep[][see Section \ref{subsec:OICIIOIII} for more discussions about the effect of the AGN]{2026arXiv260514922T}.
Further constraints on dust SED at $\lambda_{\rm rest}<60\mu{\rm m}$ are important to quantify AGN contribution to dust thermal emission \citep[][]{2023MNRAS.523.4654T}.

\subsubsection{Effect of the metallicity and sSFR in redshift evolution of $T_{\rm dust}$}\label{subsubsec:metal_sSFR}
Figure \ref{fig:Tdcorrelation} further indicates that metallicity alone is not sufficient to explain the redshift evolution of $T_{\rm dust}$, as the local DGS sample and the $z\gtrsim6$ sample share a similar $Z$ range.
The average $T_{\rm dust}$ difference between the DGS and $z\gtrsim6$ samples derived from the power-law fitting is by a factor of 2.5, suggesting another factor apart from $Z$ contributes to the redshift dependence of $T_{\rm dust}$.
If we recall that $T_{\rm dust}\propto(L_{\rm IR}/M_{\rm dust})^{1/(4+\beta_{\rm dust})}$ \citep{2022MNRAS.512...58F} and $M_{\rm dust}/M_{\rm gas}\propto Z$ \citep[e.g.,][]{2014A&A...563A..31R,2017MNRAS.471.3152P}, $T_{\rm dust}$ is proportioal to $Z^{-1/(4+\beta_{\rm dust})}$ and $(L_{\rm IR}/M_{\rm gas})^{1/(4+\beta_{\rm dust})}$.
In addition to the $Z$ dependence in Section \ref{subsec:tdustvar}, we also demonstrate $L_{\rm IR}/M_{\rm gas}$ dependence using ${\rm sSFR}$ ($\equiv{\rm SFR}/M_{\ast}$) by assuming constant stellar-to-gas mass ratio \citep[see also,][]{2019MNRAS.489.1397L}.

We find that the $z\gtrsim6$ sample with the robust stellar mass measurements (see Table \ref{tab:tab1}) has $\sim1.5\,{\rm dex}$ higher specific SFR (sSFR) than the DGS samples at the fixed metallicity, corresponding to $T_{\rm dust}$ enhancement by a factor of $\sim2$. 
We confirm the positive correlation between $T_{\rm dust}$ and sSFR in our $z\gtrsim6$ and local samples, and identify the systematic offset between these two samples likely due to the metallicity difference at the given sSFR (Appendix \ref{appendix:LIRMgas}).
Therefore, we introduce two vaiables in $T_{\rm dust}$ parametarization, $Z$ and sSFR, as $\log T_{\rm dust}=A\times(\log Z+\alpha\times\log {\rm sSFR_{\rm UV+IR}})+B$, where $\log Z$ is a unit of $12+\log({\rm O/H})$ and sSFR is based on ${\rm SFR}_{\rm UV+IR}$ for both the DGS \citep{2013PASP..125..600M,2015A&A...578A..53C,2014A&A...568A..62D} and the high-$z$ sample \citep[see][]{2024ApJ...971..161M}.
Following the procedures in the fundamental mass-metallicity ($M_{\ast}$-SFR-$Z$) relation \citep{2010MNRAS.408.2115M,2013ApJ...765..140A}, we determine the parameter $\alpha$ by minimizing the scatter around the relation.
For each value of $\alpha$, we fit $A$ and $B$ and adopt the value of $\alpha$ that yields the minimum scatter.
We find the optimal value of $\alpha=-0.56$ and obtain the following equation:
\begin{equation}
\begin{aligned}
& \log T_{\rm dust}=\\
& -0.25^{+0.03}_{-0.04}\!\times\!(\log Z\!-\!0.56\log {\rm sSFR_{\rm UV+IR}})\!+\!4.88^{+0.47}_{-0.43}.
\end{aligned}
\end{equation}
The two-variable model $T_{\rm dust}(Z,{\rm sSFR})$ yields moderately lower Bayesian information criteria (BIC) values than the single-variable models, $T_{\rm dust}(Z)$ and $T_{\rm dust}({\rm sSFR})$, with the difference in the BIC ($\Delta {\rm BIC}={\rm BIC}_{{\rm 3param}}-{\rm BIC}_{{\rm 2param}}$, \citealt{1978AnSta...6..461S}) of $-3.3$, suggesting that both $Z$ and sSFR are important parameters to explain observed $T_{\rm dust}$ across $z\sim0$ to $z\sim6$--9.
We have also tested with ${\rm SFR}_{\rm H\beta}$ 
instead of ${\rm SFR}_{\rm UV+IR}$ since both $Z$ and ${\rm SFR}_{\rm H\beta}$ can be measured from the rest-frame optical spectroscopy.
We found a similar result with $\alpha=-0.63$ as follows:
\begin{equation}
\begin{aligned}
& \log T_{\rm dust}=\\
& -0.19^{+0.03}_{-0.02}\!\times\!(\log Z\!-\!0.63\log {\rm sSFR_{\rm H\beta}})\!+\!4.19^{+0.37}_{-0.33}.
\end{aligned}
\end{equation}
These alignments strongly suggest that low-metallicity, high-sSFR galaxies exhibit high $T_{\rm dust}$ because intense radiation from star formation heats a relatively small amount of dust.

\citet{2024MNRAS.527...10V} found the potential links between $T_{\rm dust}$ and gas depretion timescale ($t_{\rm dep}$), which more directly reflects the $L_{\rm IR}/M_{\rm gas}$ than sSFR.
We also compute $L_{\rm IR}/M_{\rm gas}$ using $L_{\rm [CII]}$-to-$M_{\rm gas}$ conversion factor ($\alpha_{\rm [CII]}$) in \citet{2018MNRAS.481.1976Z} and try similar analysis as in sSFR.
We find a fitting result in $T_{\rm dust}(Z,L_{\rm IR}/M_{\rm gas})$ show around twice larger $\chi^{2}$ value than $T_{\rm dust}(Z,{\rm sSFR})$.
This is because $L_{\rm IR}/M_{\rm gas}$ is comparable between the local and $z\gtrsim6$ samples, while the sSFR of the $z\gtrsim6$ sample is $\sim1.5\,{\rm dex}$ higher than that of the local sample (see Appendix \ref{appendix:LIRMgas}).
Since it is not clear that [C\,{\sc ii}]-based $M_{\rm gas}$ estimation is reliable in the $z\gtrsim6$ sample, further constraints on the other probe of the gas budget (e.g., CO lines) will advance our understanding of the $T_{\rm dust}$ and $L_{\rm IR}/M_{\rm gas}$ (or $t_{\rm dep}$) connection.

To reproduce 2.5 times higher $T_{\rm dust}$ in the $z\gtrsim6$ sample at the given metallicity, gas column density ($N_{\rm H}$) may also be required since $N_{\rm H}$ reflects the amount of the absorbed UV light by dust, as suggested in \citet{2022MNRAS.513.3122S}.
Based on the proposed $N_{\rm H}$ dependence in \citet{2022MNRAS.513.3122S}, $T_{\rm dust}\propto N_{\rm H}^{1/6}$, $\sim5\times N_{\rm H}$ is required.
This is implied by $N_{\rm H}$ measurements from Ly$\alpha$ damping wing \citep[e.g.,][]{2024ApJ...971..124U} and higher $n_{\rm H}$ in $z\gtrsim6$ galaxies to be discussed in the following Section \ref{subsec:OICIIOIII}.
The combination of those two effects is in line with the observed offset of $T_{\rm dust}$, and is likely to result in higher $T_{\rm dust}$ at high-$z$.



%
%
%
%
%
%
\begin{figure}[t]
\begin{center}
\epsscale{1.15}
\includegraphics[width=8.5cm,bb=0 0 200 150, trim=0 1 0 0cm]{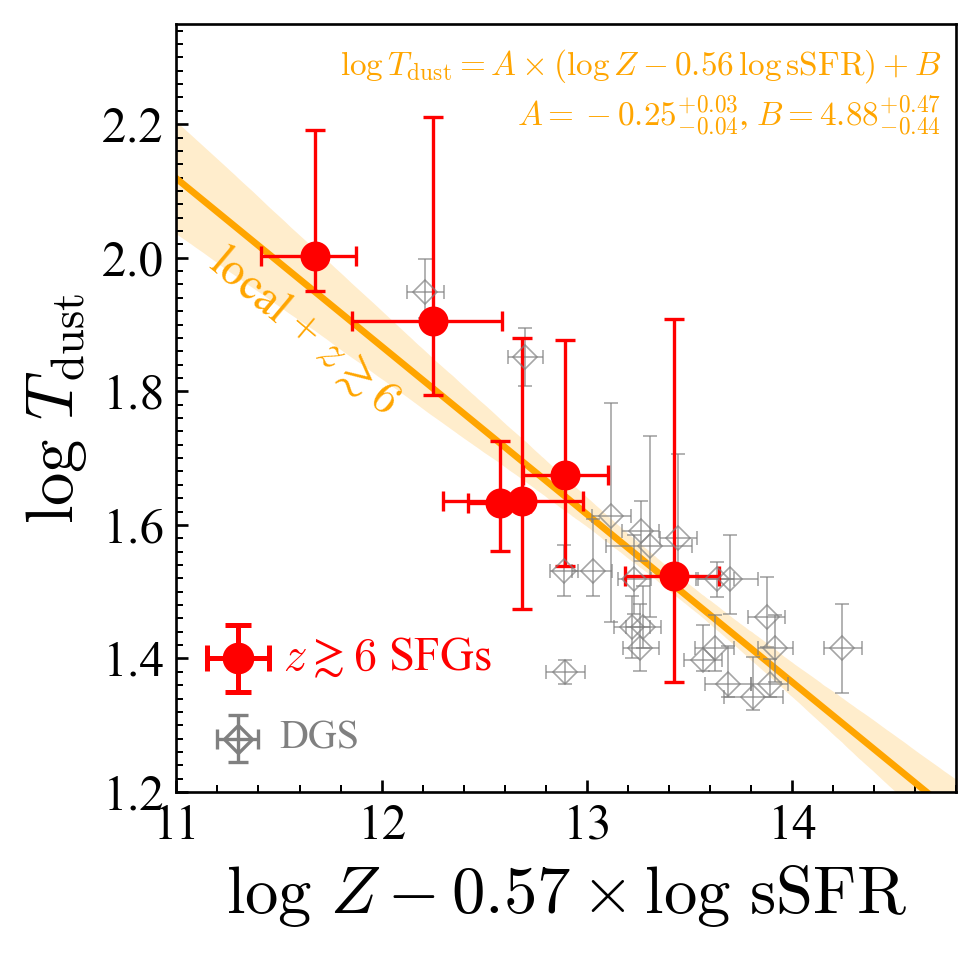}
\caption{Dust temperature as a function of metallicity and sSFR. 
The galaxies with Band-9/10 observations analysed in this work are shown as the red circles, along with the local DGS sample \citep[gray diamonds,][]{2013PASP..125..600M,2015A&A...578A..53C,2014A&A...568A..62D}.
The $T_{\rm dust}$ dependence is better explained by two variables, $Z$ and sSFR, rather than a single one, with $\log T_{\rm dust}=-0.25^{+0.03}_{-0.04}\times(\log Z-0.56\times\log {\rm sSFR_{\rm UV+IR}})+4.88^{+0.47}_{-0.43}$
}
\label{fig:Tdust_3params}
\end{center}
\end{figure}
%
%
%
%
%
%

%
%
%
%
%
%
\begin{figure*}[htbp]
\begin{center}
\epsscale{1.15}
\includegraphics[width=16.5cm,bb=0 0 1000 650, trim=0 1 0 0cm]{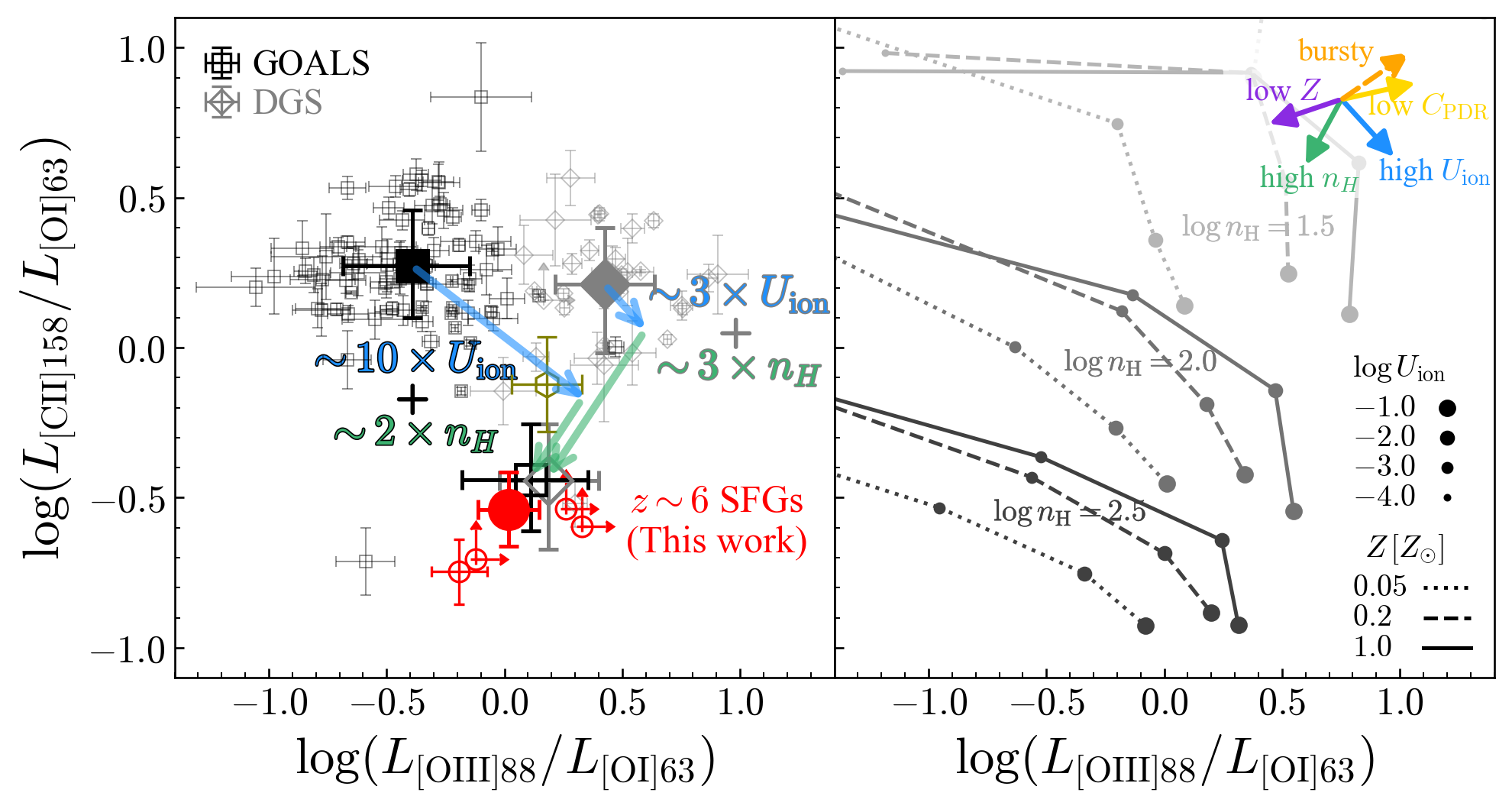}
\caption{Line ratio between \cii/\oi\ and \oiii/\oi. [left] Observational results from our individual (open) and stacked (filled) measurements at $z\gtrsim6$ are shown in the red circles, as well as those of the QSO at $z\sim6$ \citep[olive,][]{2025PASJ...77..139I} and the local samples (gray and black, DGS; \citealt{2015A&A...578A..53C}, GOALS; \citealt{2013ApJ...774...68D}). The averaged values from the DGS \citep{2015A&A...578A..53C} and the GOALS \citep{2013ApJ...774...68D} samples are shown in the large gray and black markers, respectively, as a comparison sample of the $z\gtrsim6$ galaxies. The shifts by change of $U_{\rm ion}$ and $n_{\rm H}$ from the original DGS and GOALS sample are shown in corresponding open markers. [right] Result of the \texttt{cloudy} model calculation. The light, intermediate, and dark grays correspond to densities of $\log n_{\rm H}\,[{\rm cm}^{-3}]= 0.5$, 1.0, 2.0, and 3.0.
The dotted, dashed, and solid lines are results for metallicities of $Z=0.05\,Z_{\odot}$, $0.2\,Z_{\odot}$, and $1.0\,Z_{\odot}$, respectively. The size of the marker reflects $\log U_{\rm ion}= -4.0$, -3.0, -2.0, and -1.0, respectively.
The overall effect of the parameters ($n_{\rm H}$, $U_{\rm ion}$, $Z$, $C_{\rm PDR}$, and burstiness) is illustrated by arrows in the top right.}
\label{fig:lineratio_model}
\end{center}
\end{figure*}

\subsection{ISM condition at $z\sim6$ and comparison with the local galaxies}\label{subsec:OICIIOIII}

\subsubsection{Low \cii/\oi\ ratios in $z\sim6$ galaxies}\label{subsubsec:cloudyresults}

The left panel of Figure \ref{fig:lineratio_model} shows the line ratios among \cii, \oiii, and \oi.
The stacked $z\gtrsim6$ value exhibits a \Rciioi\ that is $\sim1\,{\rm dex}$ lower than those of local samples, while the \Roiiioi\ lies between the GOALS and DGS samples.
We note that the \oi\ line can become optically thick and suffer self-absorption in dense star-forming environments \citep[e.g.,][]{1996ApJ...462L..43P,2007ApJ...662.1024A,2017ApJ...846...32D,2025PASJ...77..139I}. 
If present, such effects would modify the observed \cii/\oi\ ratio and may introduce systematic uncertainties in the inferred gas density.
While it is difficult to determine whether the absorption is present in our target galaxies, given the limited S/N in their spectra, correcting for self-absorption would increase the intrinsic \oi\ luminosity, implying an even lower intrinsic \cii/\oi\ ratio than observed.

To investigate the origin of the differences between the local and $z\gtrsim6$ samples, we perform \texttt{cloudy} calculations
using version 23.01 \citep{2023RMxAA..59..327C}.
We include both the H{\sc ii} region and the photodissociation region (PDR) to compare \oiii\ (mainly originating from the H\,{\sc ii} region) with \cii\ and \oi\ (mainly originating from the PDR; \citealt{2019A&A...626A..23C}), following \citet{2020ApJ...896...93H}.
The details of the \texttt{cloudy} set-ups are provided in \ref{appndix:cloudy_main}.

The model results are shown in the right panel of Figure \ref{fig:lineratio_model}, and the qualitative parameter dependencies are summarized in the upper-right corner of the right panel.
An increase in the ionization parameter expands the H\,{\sc ii} region while reducing the relative contribution from the neutral gas. 
The reduction is typically stronger for \cii\ than \oi, because \cii\ preferentially arises from a diffuse PDR layer that is easily suppressed in highly ionized environments, whereas \oi\ originates from denser and warmer PDR layers that are less strongly affected by the ionization parameter.
Therefore, the higher $U_{\rm ion}$ increase \oiii/\oi\ and decrease \cii/\oi\ (Figure \ref{fig:lineratio_demo} in Appendix \ref{appndix:cloudy_main}, left panels). 
Since \oi\ has a much higher critical density than either \cii\ or \oiii, increasing $n_{\rm H}$ strengthens \oi\ emission relative to \oiii\ and \cii\ because of the collisional de-excitation, resulting in lower \cii/\oi\ and \oiii/\oi\ ratios (Figure \ref{fig:lineratio_demo} in Appendix \ref{appndix:cloudy_main}, right panels).
Low metallicity generally decreases all metal-line luminosities.
However, emission lines from the PDR are less affected by metallicity, since dust shielding of FUV photons is proportional to $1/Z$, and FUV photons penetrate more deeply into the gas cloud in low-metallicity environments \citep[e.g.,][]{2006ApJ...644..283K}.
There is a small difference between \cii\ and \oi, as the expansion of \oi-emitting warm and dense gas is more than that of \cii-emitting diffuse gas.
Therefore, the higher $Z$ largely decreases \oiii/\oi\ and slightly diminishes \cii/\oi.
The low $C_{\rm PDR}$ simply decreases all the emission lines from the PDR, making the \oiii/\oi\ higher.

The lower \Rciioi\ ratio and the intermediate \Roiiioi\ ratio in $z\gtrsim6$ galaxies relative to local samples are naturally explained by higher $n_{\rm H}$ and the high $U_{\rm ion}$, given the higher critical density of \oi\ compared to \cii\ and the higher ionization potential relevant for producing \oiii\ compared to \cii\ and \oi, respectively.
The stacked $z\gtrsim6$ measurements are consistent with an approximately $\sim3\times U_{\rm ion}$ and $\sim3\times n_{\rm H}$ compared to DGS, and $\sim10\times U_{\rm ion}$ and $\sim2\times n_{\rm H}$ compared to GOALS (Figure \ref{fig:lineratio_model}).
A similar level of enhancement in $U_{\rm ion}$ and $n_{\rm H}$ is also consistent with constraints from [O\,{\sc i}] $146\,\mu{\rm m}$ observations (Appendix \ref{appendix:oi146}).
Such enhanced $U_{\rm ion}$ and $n_{\rm H}$ compared with the local galaxies has a good agreement with [O\,{\sc iii}]$\lambda5007$/[O\,{\sc ii}]$\lambda3727$ measurements \citep[e.g.,][]{2014MNRAS.442..900N} and the electron density measurements \citep[e.g.,][see also \citealt{2025ApJ...993..204H} for tracer dependence of the electron density]{2014ApJ...795..165S,2023ApJ...956..139I}.

Throughout the \texttt{cloudy} modeling in this Section, we do not consider a radiation field that is produced by AGN rather than star formation.
Since \oi\ can be a main coolant in a X-ray dominated region (XDR) produced by the X-ray radiation from AGNs \citep[][]{1996ApJ...466..561M,2009ApJ...703.1203H,2022ARA&A..60..247W}, high \oi/\cii\ ratio can be an indicator of the AGN, specifically in the local Universe \citep{2004ApJ...604..565D,2017ApJ...846...32D}.
However, the neutral ISM condition in the high-$z$ SFGs is similar to that in the AGNs given their intense star formation and high gas density \citep{2019ApJ...881...63N,2019A&A...631A.167D,2025arXiv250403831F,2025PASJ...77..139I}, as our \oi/\cii\ is reproduced by reasonable $U_{\rm ion}$ and $n_{\rm H}$ values at high-$z$ SFGs \citep[e.g.,][]{2014MNRAS.442..900N,2023ApJ...956..139I}.
For instance, \citet{2026A&A...710A.376X} compared FIR line ratios with XDR+PDR modeling in high-$z$ QSOs and found that PDRs alone cannot explain their FIR line ratio \citep[see also,][]{2021A&A...652A..66P}.
Our \oi\ detection is limited to the stacking analysis except for one tentative detection.
An individual constraint on the \oi/\cii\ and comparison with $T_{\rm dust}$ values will be helpful for further discussion about the AGN contribution in high-$z$ galaxies.

%
%
%
%
%
%
\begin{figure}[t]
\begin{center}
\epsscale{1.15}
\includegraphics[width=8.5cm,bb=0 0 200 150, trim=0 1 0 0cm]{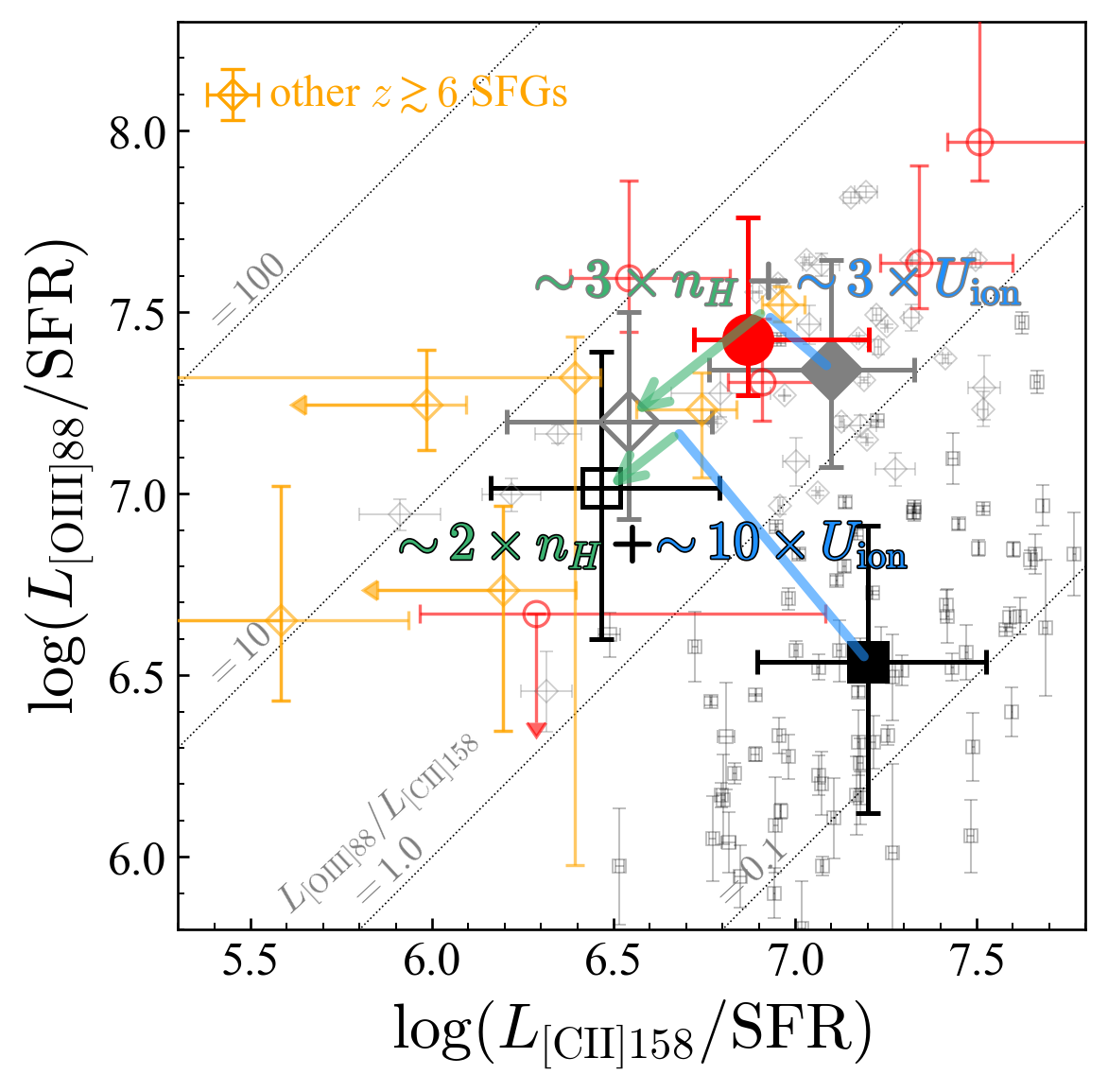}
\caption{Comparison of the \cii/SFR vs \oiii/SFR. Our individual (open) and stacked (filled) measurements at $z\gtrsim6$ are shown in the red circles. The other $z\gtrsim6$ galaxies from \citet{2020ApJ...896...93H} are also shown in the orange markers. The averaged values from the DGS \citep{2015A&A...578A..53C} and the GOALS \citep{2013ApJ...774...68D} samples are shown in the large gray and black markers, respectively, as a comparison sample of the $z\gtrsim6$ galaxies. The shifts due to changes of $U_{\rm ion}$ and $n_{\rm H}$ from the original DGS and GOALS samples are shown in the corresponding open markers as in Figure \ref{fig:lineratio_model}.
The similar levels of the $U_{\rm ion}$ and $n_{\rm H}$ enhancement from local samples in \Roiiioi-\Rciioi\ relation naturally explain the \oiii/SFR-\cii/SFR ratios at $z\gtrsim6$.}
\label{fig:lineratio_SFR}
\end{center}
\end{figure}

\subsubsection{Origin of high \oiii/\cii ratio at $z\gtrsim6$}\label{subsec:highOIIICII}

We next place the \oi, \cii, and \oiii\ line ratios in the broader context of the high \Roiiicii\ ratios observed at $z\gtrsim6$.
We derive $L_{\rm [CII]158}$/SFR and $L_{\rm [OIII]88}$/SFR for both individual and stacked results, and compare \texttt{cloudy} calculations as well as the \oiii, \cii, and \oi\ line ratios.
We compute the total SFR of the $z\gtrsim6$ sample as ${\rm SFR}_{\rm UV}+{\rm SFR}_{\rm IR}$, adopting the conversion factors from \citet{2014ARA&A..52..415M} with \citet{2003PASP..115..763C} IMF.
%
%
We overplot the expected shifts in this plane driven by changes in $U_{\rm ion}$ and $n_{\rm H}$ based on the \texttt{cloudy} calculations.
Here, SFR is computed from H$\alpha$ luminosity using the conversion factor in \citet{1998ARA&A..36..189K}, and is rescaled to values in the \citet{2003PASP..115..763C} IMF following \citet{2020ApJ...896...93H}. 
While we adopt a 1\,Myr instantaneous burst in fiducial \texttt{fsps} models as described in Section \ref{subsec:OICIIOIII} and Appendix \ref{appndix:cloudy_main}, we confirm that the results do not change if we apply 100\,Myr constant star formation \citep[see][for more discussion about the conversion factors]{2022ApJ...935..119S}.

The results are shown in Figure \ref{fig:lineratio_SFR}. 
Higher $U_{\rm ion}$ increases \Roiiicii, while higher $n_{\rm H}$ suppresses both \cii/SFR and \oiii/SFR, moving local averages into the regime occupied by our $z\gtrsim6$ sample and the other $z\gtrsim6$ galaxies in \citet{2020ApJ...896...93H} on the \cii/SFR–\oiii/SFR plane.
Another factor that enhance \Roiiicii\ is small PDR covering fraction ($C_{\rm PDR}$) as discussed in \citet{2020ApJ...896...93H}.
\oiii/\oi\ is inversely proportional to $C_{\rm PDR}$ and \cii/\oi\ is almost independent on $C_{\rm PDR}$ assuming $L_{\rm [CII]158, PDR}/L_{\rm [CII]158}=0.9$ as shown in right panel of Figure \ref{fig:lineratio_model}.
As $z\sim6$ galaxies do not show higher \oiii/\oi\ compared with the local samples, small $C_{\rm PDR}$ is not likely to be the main reason for high \Roiiicii\ at $z\gtrsim6$.
These results suggest that the elevated $U_{\rm ion}$ and $n_{\rm H}$ provide a natural explanation for the high \Roiiicii\ ratios at $z\gtrsim6$.

Indeed, our $z\sim6$ galaxies utilized in this work are relatively massive \citep[$\log M_{\ast}/M_{\odot}\sim10.5$,][]{2020ApJ...896...93H} compared with some extremely high \Roiiicii\ galaxies $z\gtrsim6$ galaxies like MACS0416-JD1 or SXDF-NB1006-2 \citep[$\log M_{\ast}/M_{\odot}\sim8$,][]{2024MNRAS.533.2488M,2025MNRAS.544.4722R}.
As shown in Figure \ref{fig:lineratio_SFR}, such high \Roiiicii\ galaxies lie outside the shifted local averages with high $U_{\rm ion}$ and $n_{\rm H}$, variations in $C_{\rm PDR}$ may be important at low-mass galaxies with extremely high \Roiiicii\ values \citep[see also,][]{2020ApJ...896...93H,2025ApJ...990...29H}.
Further observations of the emission lines with different ionization potential and critical density from \oiii\ and \cii\ for such a high \oiii-to-\cii\ ratio object ($\text{\Roiiicii}\gtrsim10$) will be key to fully understanding the role of $C_{\rm PDR}$.

We compare our inferred high $U_{\rm ion}$ and high $n_{\rm H}$ with previous studies. \citet{2020ApJ...896...93H} argued that a combination of $\sim10\times U_{\rm ion}$ and/or $\sim0.1\times C_{\rm PDR}$ can naturally reproduce the observed \Roiiicii\ values at $z\gtrsim6$ in the \cii/SFR–\oiii/SFR plane \citep[see also,][]{2022ApJ...935..119S}.
They also discuss the effect of $n_{\rm H}$ on \Roiiicii, and indeed, $n_{\rm H}$ has a similar effect to $C_{\rm PDR}$ in the \cii/SFR-\oiii/SFR plane.
While our results are broadly consistent with \citet{2020ApJ...896...93H}, our results suggest that gas density likely plays a more important role than the PDR covering fraction from new constraints on the dense PDR tracer, i.e., \oi\ line luminosities.

Using zoom-in simulations, \citet{2025A&A...704A..39K} investigated correlations between \Roiiicii\ and several physical properties, including $U_{\rm ion}$, $Z$, $n_{\rm H}$, and burstiness. They found that \Roiiicii\ is enhanced for $\log U_{\rm ion}>-1.5$, $\log n_{\rm H}\,[{\rm cm}^{-3}]\sim2.5$, and merger-driven starburst activity, in good agreement with our inferred conditions.
\citet{2025arXiv250512397N} further suggested that the mass fraction of ionized versus neutral gas may correlate with \Roiiicii, likely as a consequence of high $U_{\rm ion}$.

As suggested in \citet{2021MNRAS.505.5543V}, burstiness of the star formation possibly correlate to (surface) \Roiiicii\ ratio.
More recently, \citet{2025arXiv250916071A} argued that burstiness, rather than nebular parameters such as $U_{\rm ion}$ and $Z$, is the primary driver of high \Roiiicii\ at $z\gtrsim6$. 
Since \oiii\ originates in H\,{\sc ii} regions and traces newly formed stars, whereas \cii\ largely arises from more diffuse gas farther from star-forming regions, \oiii\ can be enhanced relative to \cii\ during recent bursts of star formation \citep[see][]{2025arXiv250512397N}.
To test the potential effect of the burstiness, we additionally run \texttt{cloudy} calculations with basically the same parameter sets as in Section \ref{subsec:OICIIOIII}.
We virtually reproduce the burstiness by combining the underlying continuous star-forming component with the instantaneous burst component at different burst ages and relative burst strengths (see Appendix \ref{appendix:bursty} for details).

The impact of burstiness is shown on the orange arrow in Figure \ref{fig:lineratio_model}.
Increasing burstiness changes the stellar SED shape toward a younger, massive star-dominated population, simultaneously enhancing both the ionizing and FUV radiation fields. 
The harder and younger stellar population simply boosts the highly ionized \oiii\ emission, leading to higher \oiii/\oi\ ratios. 
At the same time, the enhanced FUV field more efficiently boosts the cooling from the more diffuse C$^+$-emitting layer than from the warmer and denser neutral gas traced by \oi. 
As a result, \cii\ increases more strongly than \oi, producing elevated \cii/\oi ratios toward more bursty models.
Since an increase in burstiness enhances both \oiii/\oi\ and \cii/\oi, increased burstiness does not directly align with a low \cii/\oi\ ratio and may not be the primary driver of the high \oiii/\cii\ ratios at $z\gtrsim6$ galaxies.

The sample dependence is also crucial to consider.
Current \oiii/\cii\ observed samples at $z\gtrsim6$ may be strongly biased toward UV-bright, highly star-forming, and likely bursty systems \citep[e.g.,][]{2025arXiv250916071A}. 
The elevated \oiii/\cii\ ratios may not reflect a systematic offset between $z\sim0$ and $z\gtrsim6$ galaxy populations, but instead arise because observations preferentially select galaxies caught in bursty phases \citep[e.g.,][]{2025ApJ...985..126G,2026arXiv260116284M}. 
In this case, no systematic differences in physical conditions such as $U_{\rm ion}$ or $n_{\rm H}$ would necessarily be required.
On the other hand, if \oiii/\cii\ is systematically elevated at fixed galaxy properties or across the overall galaxy population at high redshift \citep[e.g.,][]{2026A&A...707A...5M}, then burstiness alone would likely be insufficient to explain the trend. 
In that case, systematic offsets in ISM conditions, such as higher ionization parameters or different gas densities, would be required.
Interestingly, our current sample includes galaxies with enhanced \oiii/\cii\ ratios that do not appear to be strongly bursty (e.g., J1211-0118 with $\log\,EW_{\rm [OIII]+H\beta}\sim2.1$ and $\log\,\text{\Roiiicii}\sim0.6$, \citealp{2025ApJ...993..204H}), which are offset from the $EW_{\rm [OIII]+H\beta}$-\Roiiicii\ relation in \citet{2025arXiv250916071A}. 
This result favors the latter scenario, in which systematic differences in ISM conditions play a significant role.

Additionally, $EW_{{\rm [OIII]+H}\beta}$ is used as a proxy for burstiness in \citet{2025arXiv250916071A}; however, separating burstiness from nebular conditions (e.g., $U_{\rm ion}$ and electron density $n_e$) is challenging, as also noted by \citet{2025arXiv250916071A}. 
Further investigations of burstiness in low-\Roiiicii\ galaxies \citep[e.g.,][]{2024MNRAS.532.2270B} will help to assess the role of burstiness robustly.

Finally, a low carbon-to-oxygen abundance ratio (C/O) may contribute to enhanced \Roiiicii\ \citep{2025A&A...702A.260N}, given that C/O abundance varies with metallicity \citep[e.g.,][]{2017MNRAS.466.4403N}. However, the metallicity ranges of the $z\gtrsim6$ sample and the local DGS sample are similar, suggesting that C/O variations are likely subdominant in our case \citep[see also][]{2025arXiv250916071A}.


%
%
%
%
%
%
\section{Summary and Conclusions}\label{sec:summary}
In this paper, we have examined the dust continuum emissions at the rest-frame $63\,\mu{\rm m}$ of nine SFGs at $z\sim5.8$--8.3 and simultaneously observed [O\,{\sc i}]$63\,\mu{\rm m}$ emission lines of five SFGs at $z\sim6$ by utilizing ALMA's high-frequency band observations.
By applying uniform analysis towards both our main sample and supplemental sample, we measure $T_{\rm dust}$ and $M_{\rm dust}$ by the rest-frame short-wavelength constraints from the MBB fitting and obtain \oi\ line strength via individual and stacking analysis.
The rest-frame $63\,\mu{\rm m}$ dust continuum coverage is critical to constrain $T_{\rm dust}$, specifically to determine whether galaxies have $T_{\rm dust}\gtrsim60\,{\rm K}$ or $\lesssim60\,{\rm K}$.
\oi\ strength relative to \cii\ and \oiii\ provides us with constraints about key ISM parameters, such as the ionization parameter $U_{\rm ion}$, metallicity $Z$, and gas density $n_{\rm H}$.

We confirmed that $T_{\rm dust}$ of the galaxies at $z\sim6$--9 is generally $\sim30$--$60\,{\rm K}$, whereas several galaxies exhibits significantly high $T_{\rm dust}$ ($\gtrsim60\,{\rm K}$).
We explored the correlation between $T_{\rm dust}$ and several parameters: the \oiii-to-\cii\ luminosity ratio \Roiiicii, dust continuum size $r_{\rm dust}$, and gas-phase metallicity $Z$.
We found no correlation between $T_{\rm dust}$ and \Roiiicii\ or $r_{\rm dust}$, but a potential correlation with $Z$, although the statistics are insufficient.
The power-law slope of the $T_{\rm dust}$-$Z$ relation at $z\gtrsim6$ is $-0.50\pm0.19$, which is comparable with those for the local DGS sample ($-0.46\pm0.06$) if the correlation truly exists.
The dependence of the $T_{\rm dust}$ on $Z$ is larger than the analytical expectation ($T_{\rm dust}\propto Z^{-1/6}$), potentially because of the compact star-forming region due to the inefficient metal cooling, or high effective temperature in the metal-poor environment.

Based on the systematic offset of $T_{\rm dust}$ at the fixed metallicity between $z\sim0$ and $z\sim6$--9, we introduced additional parameters, sSFR, to understand $T_{\rm dust}$ dependence uniformly.
We find that the two-variable model $T_{\rm dust}(Z, {\rm sSFR})$ better represents the dependence of $T_{\rm dust}$ statistically compared to the single-variable models ($T_{\rm dust}(Z)$ or $T_{\rm dust}({\rm sSFR})$) with $\log T_{\rm dust}=-0.250^{+0.030}_{-0.032}\times(\log Z-0.57\times\log {\rm sSFR_{\rm UV+IR}})+4.895^{+0.431}_{-0.405}$, suggesting $T_{\rm dust}$ is well described by $Z$ and sSFR.

From the stacked detection of the \oi\ line, we found low \cii/\oi\ ratios in $z\sim6$ galaxies.
As \oi\ has a higher critical density than the \cii\ line, low \cii/\oi\ ratios imply that \cii\ is collisionally de-excited due to the high gas density.
With a \texttt{cloudy} model calculation, we infer the $\log n_{\rm H}\,[{\rm cm}^{-3}]\sim2.5$ and $\log U_{\rm ion}\sim-2.0$ at $z\sim6$ galaxies on average, which is $\sim3$--$10\times$ higher $U_{\rm ion}$ and $\sim2$--$3\times$ higher $n_{\rm H}$ from local samples.
The enhanced $U_{\rm ion}$ and $n_{\rm H}$ nicely reproduce the \oiii/SFR-\cii/SFR relation.
We suggest that the enhancement of not only $U_{\rm ion}$ but also $n_{\rm H}$ is crucial for explaining the high \Roiiicii\ at $z\gtrsim6$, while several other possible origins, such as burstiness and a high C/O ratio, may also be important.

Our study demonstrated the unique capability of the ALMA high-frequency bands to constrain $T_{\rm dust}$ for high-$z$ galaxies.
Furthermore, the combination of ALMA and JWST demonstrated the potential connection between FIR dust emission and ISM properties.
Improvement of the statistics with a large sample is critical to confirm (or rule out) the correlation between $T_{\rm dust}$ and other physical parameters, such as $Z$.
For galaxies with $T_{\rm dust}\gtrsim60\,{\rm K}$, Band-9 observations are still at too long a rest-frame wavelength to capture the peak of the FIR SED.
Further short-wavelength observations using Band-10 will allow us to better constrain $T_{\rm dust}$ and are crucial to unveil the origin of their high $T_{\rm dust}$.

\oi\ is a key probe of the warm, dense neutral gas in the ISM. However, at $z\lesssim6$, the \oi\ line falls in high-frequency ALMA bands (i.e., Band-9 or 10), making observations challenging. 
At higher redshifts, the \oi\ line shifts to lower frequencies, including Band 8 and below, making it a promising tracer for investigating the ISM conditions of high-redshift galaxies.

\acknowledgments
We thank Yurina Nakazato, Katherine A. Suess, Ben Forrest, and Olivia R. Cooper for giving us helpful comments. 
This paper makes use of the following ALMA data: ADS/JAO.ALMA\#2015.1.00540.S, \#2015.1.01406.S, \#2016.1.00954.S, \#2017.1.00190.S, \#2017.1.00508.S, \#2017.1.00697.S, \#2017.1.00775.S, \#2019.1.01634.L, \#2021.1.01297.S, \#2021.1.00318.S, \#2022.1.00522.S, \#2022.1.01324.S, \#2023.1.00022.S, \#2023.1.00629.S, \#2023.1.01033.S, and \#2024.1.00537.S. 
ALMA is a partnership of ESO (representing its member states), NSF (USA), and NINS (Japan), together with NRC (Canada), MOST and ASIAA (Taiwan), and KASI (Republic of Korea), in cooperation with the Republic of Chile. The Joint ALMA Observatory is operated by ESO, AUI/NRAO, and NAOJ.
Data analysis was in part carried out on the Multi-wavelength Data Analysis System operated by the Astronomy Data Center (ADC), National Astronomical Observatory of Japan.
I.M. acknowledges funding from JWST-GO-04111.035.
Y.S. is supported by JSPS KAKENHI Grant Number JP26K17200.
KK acknowledges the support by JSPS KAKENHI Grant Numbers JP22H04939, JP23K20035, and JP24H00004.

\appendix

\section{Comparison of the galaxies in this work and other high-$z$ galaxies}\label{appendix:sample_bias}

As described in Section \ref{subsec:main}, the main sample is composed of the Band-9 follow-ups towards the original SERENADE sample.
The follow-up target is selected from Band-6 and 8 observations covering rest-frame 88\,$\mu$m and 158\,$\mu$m.
To check potential bias in the average $T_{\rm dust}$ from stacking analysis ($\sim40\,{\rm K}$) and the fraction of the galaxy showing $T_{\rm dust}\gtrsim60\,{\rm K}$ (3/9, Figure \ref{fig:zTdust}), we compare the flux ratio between rest-frame 88\,$\mu$m and 158\,$\mu$m in our sample to the other $z\gtrsim6$ galaxies with dust continuum detection either at 88\,$\mu$m or 158\,$\mu$m ($S_{\rm 88\,\mu{\rm m}}/S_{\rm 158\,\mu{\rm m}}$).

\begin{figure}[t]
\begin{center}
\epsscale{1.15}
\includegraphics[width=8.5cm,bb=0 0 200 150, trim=0 1 0 0cm]{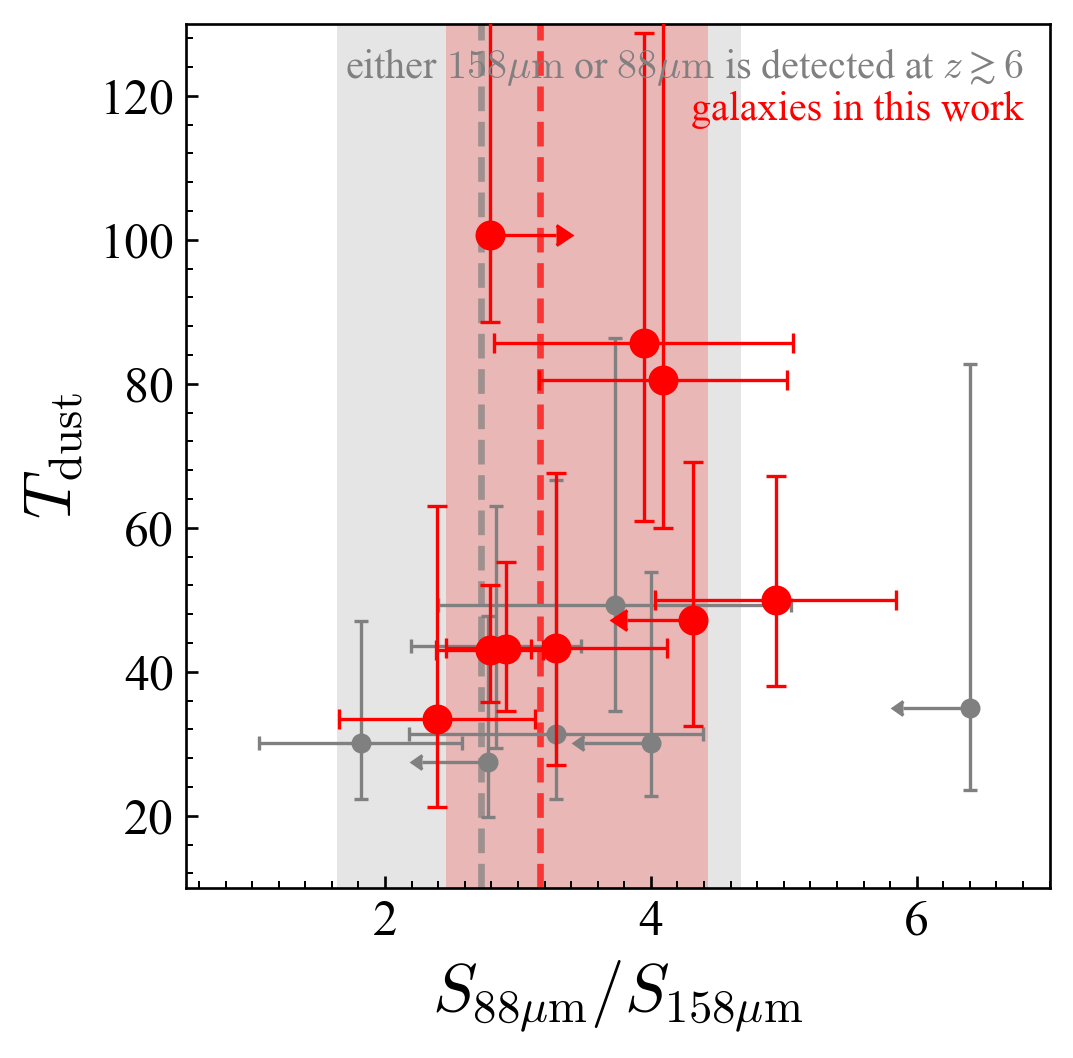}
\caption{Dust temperature as a function of the continuum flux ratio between rest-frame 88\,$\mu$m and 158\,$\mu$m.
The galaxies analyzed in this work and the other $z\gtrsim6$ galaxies with dust continuum detection either at 88\,$\mu$m or 158\,$\mu$m \citet{2024ApJ...971..161M} are shown in red and gray markers, respectively. 
Since the flux ratio in this work and other galaxies at $z\gtrsim6$ do not show a significant difference, the sample in this work is not likely to be strongly biased to the high $T_{\rm dust}$ objects.}
\label{fig:sample_bias}
\end{center}
\end{figure}

Figure \ref{fig:sample_bias} show $T_{\rm dust}$ as a fuction of $S_{\rm 88\,\mu{\rm m}}/S_{\rm 158\,\mu{\rm m}}$.
We find a slightly higher $S_{\rm 88\,\mu{\rm m}}/S_{\rm 158\,\mu{\rm m}}$ for the sample in this work compared with that of the other galaxies at $z\gtrsim6$.
However, the difference is not significant, and this implies that there is no strong sample bias in the sample in this work.

%
%
%
%
%
%

\section{Thumnails and MCMC fitting results for the supplemental sample}\label{appendix:Tdfitting_supp}

\begin{figure}[t]
\begin{center}
\epsscale{1.15}
\includegraphics[width=8.5cm,bb=0 0 200 150, trim=0 1 0 0cm]{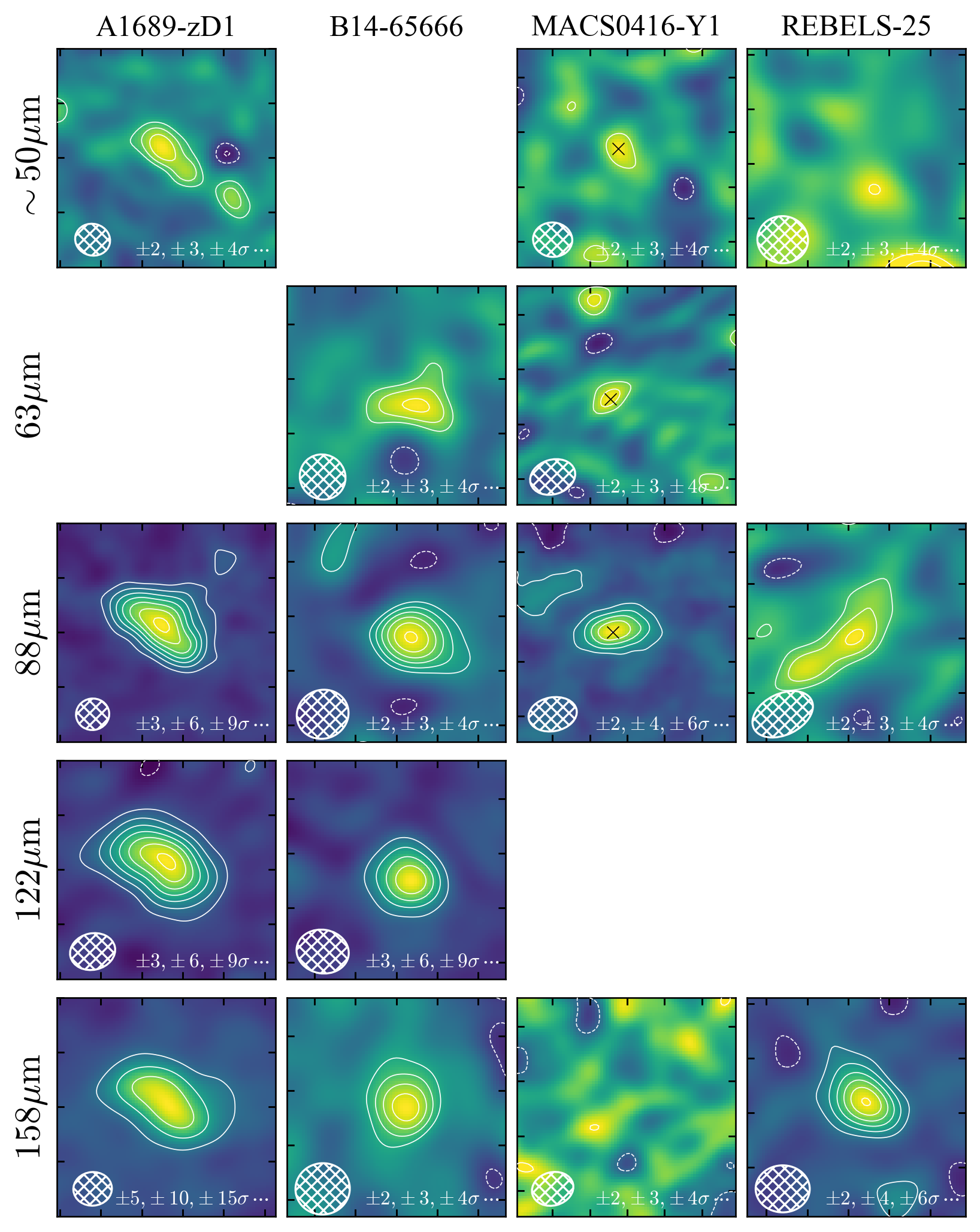}
\caption{Same with Figure \ref{fig:thumnail_dust}, but for the supplemental ($T_{\rm dust}$) sample. Here, short wavelength observations in $\sim50\,\mu{\rm m}$ cover slightly different rest-frame wavelength \citep[see Table \ref{tab:tab2} and][]{2021MNRAS.508L..58B,2025MNRAS.544.1502B}.}
\label{fig:thumnail_dust_supp}
\end{center}
\end{figure}

\begin{figure*}[!htbp]
\begin{center}
\centering
\includegraphics[width=0.95\textwidth]{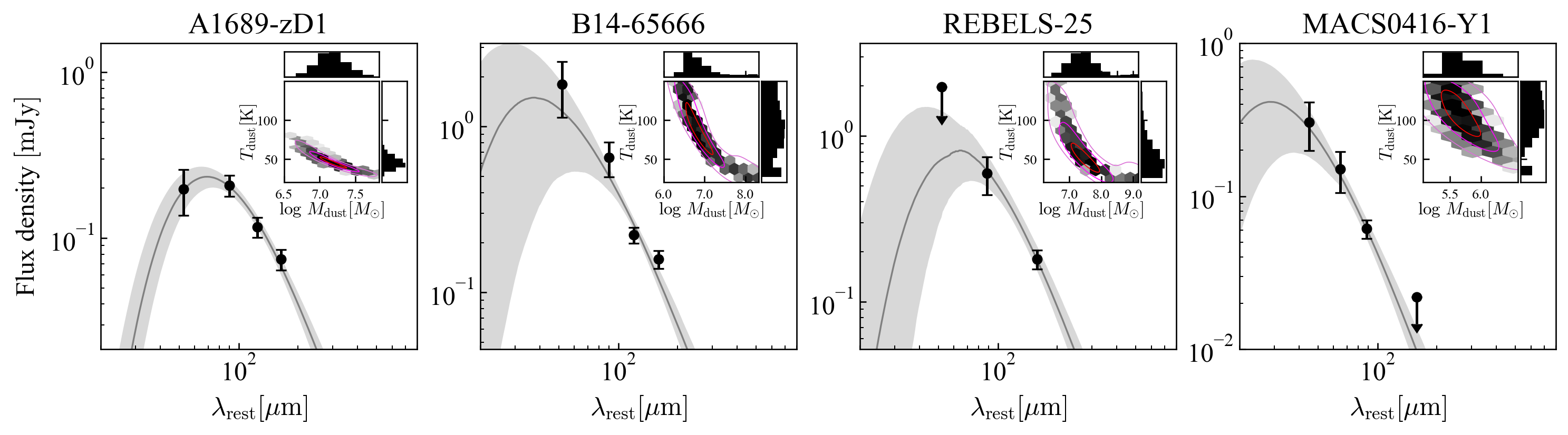}
\caption{Same as Figure \ref{fig:Tdfitting}, but for the supplemental ($T_{\rm dust}$) sample \citep[see also,][]{2015Natur.519..327W,2017MNRAS.466..138K,2019PASJ...71...71H,2020MNRAS.495.1577I,2021ApJ...923....5S,2021MNRAS.508L..58B,2022MNRAS.515.1751W,2022ApJ...934...64A,2023MNRAS.518.6142A}.}
\label{fig:Tdfitting_supp}
\end{center}
\end{figure*}

In sections \ref{subsec:flux} and \ref{subsec:mcmc}, we re-analyze archival data for the supplemental ($T_{\rm dust}$) sample.
Basically, we follow the same methodology as that used in the main+supplemental ($T_{\rm dust}$, \oi) samples.
The continuum images and results of the MCMC fitting are shown in Figure \ref{fig:thumnail_dust_supp} and \ref{fig:Tdfitting_supp}, respectively.

%
%
%
%
%
%
%
%
%
%
%
%

%
%
%
%
%
%

\section{Comparison between $L_{\rm IR}/M_{\rm gas}$ and specific SFR}\label{appendix:LIRMgas}

\begin{figure*}[!htbp]
\begin{center}
\centering
\includegraphics[width=0.95\textwidth]{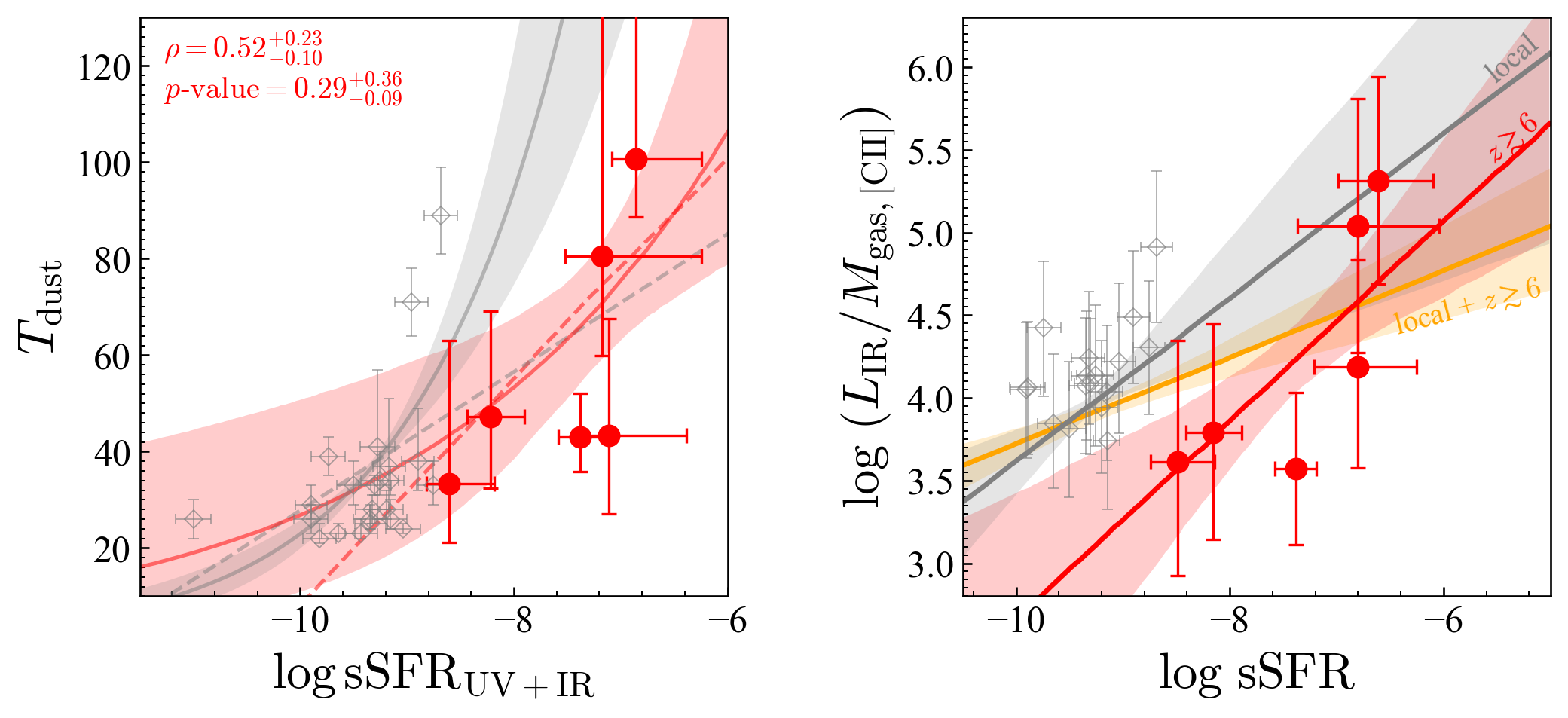}
\caption{[left] Dust temperature as a function of sSFR. The markers and lines are the same as the right panel of Figure \ref{fig:Tdcorrelation}. [right] Comparison between sSFR and $L_{\rm IR}/M_{\rm gas}$ for the local DGS sample (gray) and the $z\gtrsim6$ sample (red).
While local, $z\gtrsim6$ and combined samples show tentative positive correlation, the person's correlation indicates that the correlation is not statistically significant ($p$-${\rm values}>0.1$).}
\label{fig:LIRMgas_comp}
\end{center}
\end{figure*}

As computed in Section \ref{subsubsec:metal_sSFR}, $T_{\rm dust}$ is supposed to correlate with $L_{\rm IR}/M_{\rm dust}$ in addition to $Z$. 
We mainly utilize sSFR instead of $L_{\rm IR}/M_{\rm dust}$ in Section \ref{subsubsec:metal_sSFR}, owing to difficulties in measuring gas masses at high-$z$.
The left panel of Figure \ref{fig:LIRMgas_comp} shows the $T_{\rm dust}$ as a function of sSFR.
We find a positive correlation between $T_{\rm dust}$ and sSFR in both the local and $z\gtrsim6$ samples.
As in the $T_{\rm dust}$ and metallicity correlation shown in the right panel of Figure \ref{fig:Tdcorrelation}, we also identify a systematic offset between the local and $z\gtrsim6$ samples, plausibly due to the different metallicity at the given sSFR.

Here we further test a correlation between $L_{\rm IR}/M_{\rm gas}$ and sSFR using [C\,{\sc ii}] luminosity as a gas mass tracer.
We use $L_{\rm [CII]}$-to-$M_{\rm gas}$ conversion factor of $\alpha_{\rm [CII]}=31$ and systematic 0.2\,dex uncertainty following \citet{2018MNRAS.481.1976Z}.
Figure \ref{fig:LIRMgas_comp} shows the result of the comparison between sSFR and $L_{\rm IR}/M_{\rm gas}$.
Although they show a tentative positive correlation, $L_{\rm IR}/M_{\rm gas}$ is comparable between the local and $z\gtrsim6$ samples, despite their $\sim1.5\,{\rm dex}$ difference in sSFR.
This may be due to uncertainties in the gas mass estimate based on [C\,{\sc ii}].

%
%
%
%
%
%

\section{Details of the cloudy calculation and results}\label{appndix:cloudy_main}

To help interpret our \oi\ observations, we compute line ratios of \oi\, \oiii\, and \cii\ under several physical conditions using \texttt{cloudy} broadly following \citet{2020ApJ...896...93H}.
Under a pressure-equilibrium gas cloud with a plane-parallel geometry, we vary three primary parameters: the hydrogen density at the ionization front ($n_{\rm H}$), gas-phase metallicity ($Z$), and ionization parameter ($U_{\rm ion}$). 
Specifically, we explore $\log n_{\rm H}\,[{\rm cm}^{-3}]=0.5$–3.0 in steps of 0.5, $\log U_{\rm ion}=-4.0$ to $-0.5$ in steps of 0.5, and $Z\,[Z_{\odot}]=0.05$, 0.2, and 1.0.
As the input spectrum, we adopt BPASS v2.2 \citep{2018MNRAS.479...75S} assuming an instantaneous burst with an age of 1\,Myr, and \texttt{135\_100} IMF for binary populations (IMF slope of $-1.35$ at $0.1$--$1.0\,M_{\odot}$ and $-2.35$ at $1.0$--$100\,M_{\odot}$).
We assume solar elemental abundances and include Orion-type graphite and silicate grains. 
Calculations are stopped at $A_V=100,{\rm mag}$ to encompass the full \cii/\oi-emitting region, following \citet{2005ApJS..161...65A}.

We introduce an additional parameter following \citet{2020ApJ...896...93H}, the PDR covering fraction ($C_{\rm PDR}$), defined as the fraction of sightlines covered by the PDR.
Given that \oi\ originates from PDR and \cii\ can arise from both PDR and H\,{\sc ii} region, both \Rciioi\ and \Roiiioi\ depend on $C_{\rm PDR}$.
However, $\gtrsim90\%$ of the \cii\ emission arises from the PDR ($L_{\rm [CII]158, PDR}/L_{\rm [CII]158}\gtrsim0.9$, \citealt{2019A&A...626A..23C,2025arXiv250403831F}), and therefore the dependence of \Rciioi\ on $C_{\rm PDR}$ is weak.
We note that our conclusions are unchanged for $L_{\rm [CII]158, PDR}/L_{\rm [CII]158}\gtrsim0.5$.
The stacked \Rciioi\ and \Roiiioi\ ratio is consistent with the $\log n_{\rm H}\,[{\rm cm}^{-3}]\sim2.5$ and $\log U_{\rm ion}\sim-2.0$ under their metallicity range ($Z\sim0.2$--1.0\,$Z_{\odot}$) if $C_{\rm PDR}\sim1.0$.
There is no significant difference if we assume $C_{\rm PDR}\sim0.5$, although inferred $n_{\rm H}$ and $U_{\rm ion}$ become slightly higher and lower, respectively, than the case of $C_{\rm PDR}\sim1.0$.

%
%
%
%
%
%
\begin{figure*}[htbp]
\begin{center}
\epsscale{1.15}
\includegraphics[width=15cm,bb=0 0 1000 650, trim=0 1 0 0cm]{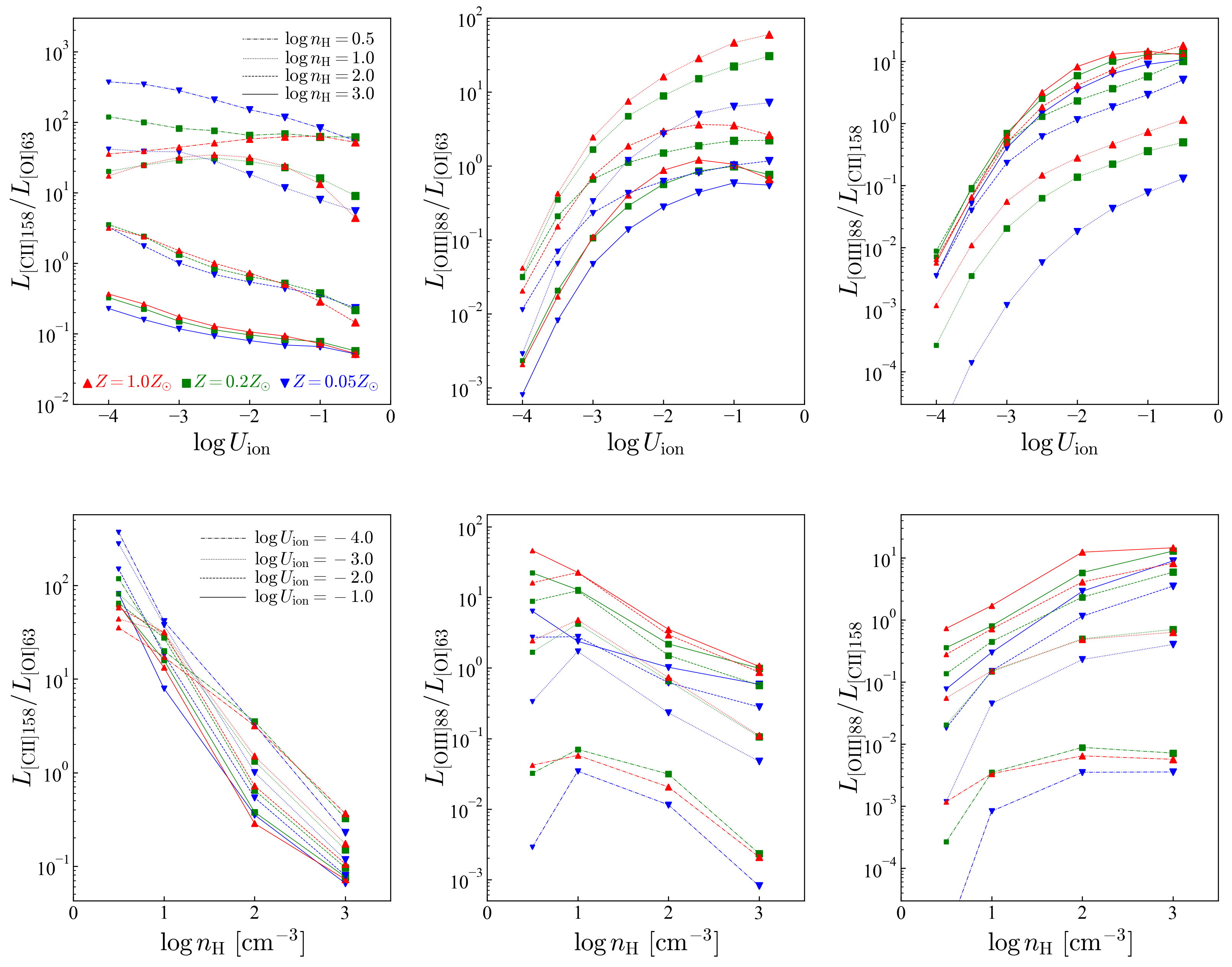}
\caption{Line ratios (\cii/\oi\ and \oiii/\oi) as a function of the $U_{\rm ion}$ (two left panels) or $n_{\rm H}$ (two right panels). The blue, green, and red lines are results for metallicities of $Z=0.05\,Z_{\odot}$, $0.2\,Z_{\odot}$, and $1.0\,Z_{\odot}$, respectively. The dotted, dashed, and solid lines correspond to densities of $\log n_{\rm H}\,[{\rm cm}^{-3}]= 0.5$, 1.0, 2.0, and 3.0, respectively, in two left panels, and $\log U_{\rm ion}= -4.0$, -3.0, -2.0, and -1.0, respectively, in two right panels. }
\label{fig:lineratio_demo}
\end{center}
\end{figure*}

\section{Comparison in [O\,{\sc i}]\,$146\,\mu\text{\lowercase{m}}$}\label{appendix:oi146}
\begin{figure}[htbp]
\begin{center}
\epsscale{1.15}
\includegraphics[width=8.5cm,bb=0 0 200 150, trim=0 1 0 0cm]{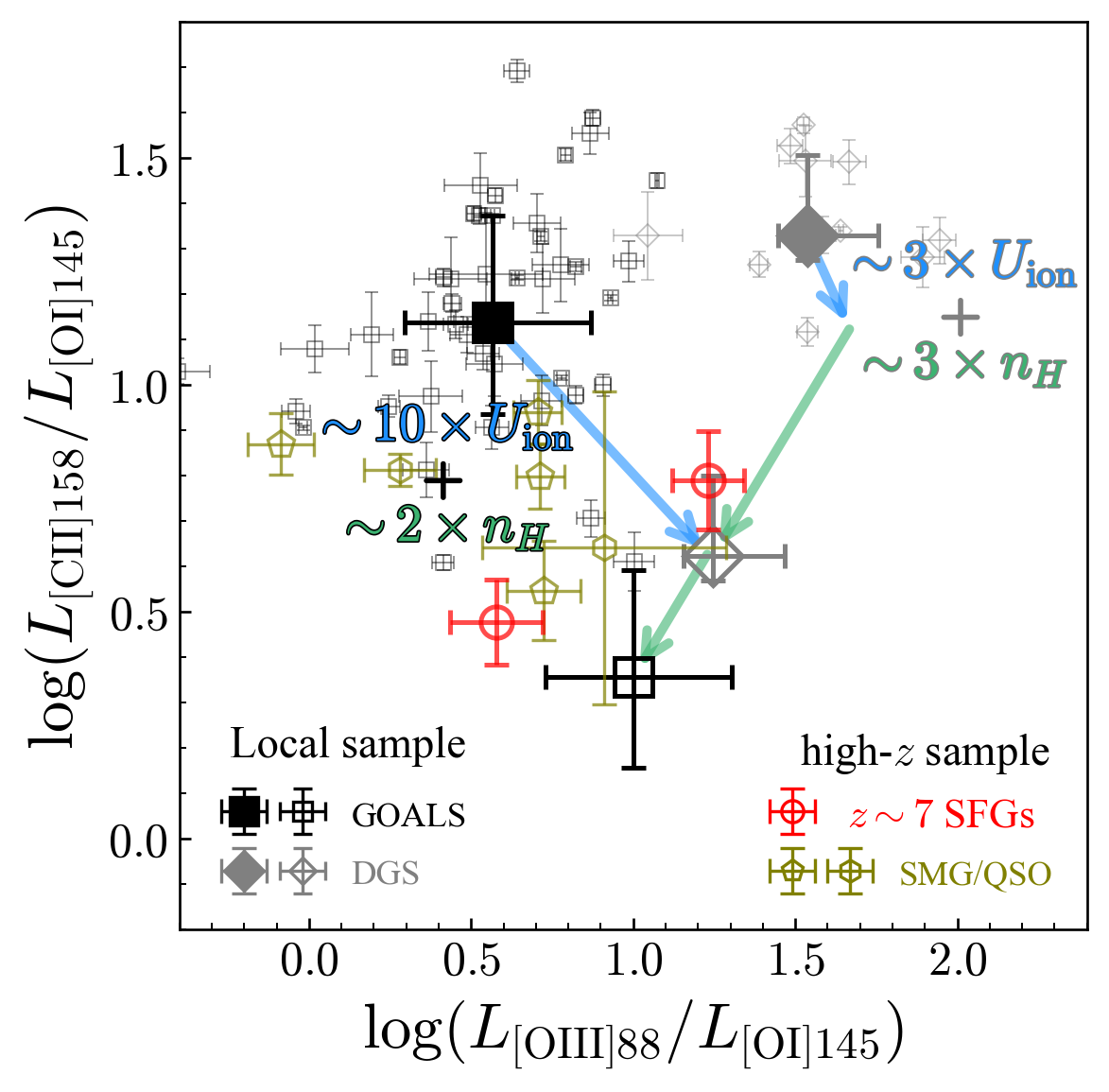}
\caption{Same with the Figure \ref{fig:lineratio_SFR}, but for [O\,{\sc i}]\,$146\mu{\rm m}$. 
We plot the results from previous studies for $z\sim7$ SFGs \citep[red circles,][]{2025arXiv250403831F}, $z\sim4$--7 SMGs \citep[olive pentagons,][]{2019A&A...631A.167D,2021ApJ...913...41L,2022ApJ...928..179L,2023ApJ...949...87L}, $z\sim4$--8 QSOs \citep[olive hexagons,][]{2019ApJ...881...63N,2020ApJ...900..131L,2021ApJ...913...41L,2022ApJ...927..152M}, and the local samples (DGS; \citealt{2015A&A...578A..53C} in the gray diamonds, GOALS; \citealt{2013ApJ...774...68D} in the black squares).
The shifts of the local samples based on the enhanced $U_{\rm ion}$ and $n_{\rm H}$ are shown following the Figure \ref{fig:lineratio_SFR}.}
\label{fig:lineratio_app}
\end{center}
\end{figure}

Since the [O\,{\sc i}] $63\,\mu{\rm m}$ becomes optically-thick under $A_V>1$ and is self-absorbed \citep{1985ApJ...291..722T}, we have also made a comparison in [O\,{\sc i}] $146\,\mu{\rm m}$ with \texttt{cloudy} models instead of [O\,{\sc i}] $63\,\mu{\rm m}$.
In Figure \ref{fig:lineratio_app}, we plot the [O\,{\sc i}] $146\,\mu{\rm m}$ luminosities against \oiii\ and \cii\ luminosities as in Figure \ref{fig:lineratio_model}, for a sample presented in \citet{2025arXiv250403831F} and several high-$z$ SMGs/QSOs \citep{2019ApJ...881...63N,2020ApJ...900..131L,2021ApJ...913...41L,2022ApJ...927..152M}.
We implement a similar enhancement of the $n_{\rm H}$ and $U_{\rm ion}$ to the local samples as in Figure \ref{fig:lineratio_model}, and compare their distribution with that of $z\gtrsim4$ galaxies.
Overall, a similar degree of the $n_{\rm H}$ and $U_{\rm ion}$ enhancements as in \oi\ explain well the $z\gtrsim4$ galaxies, and therefore, we found a consistent result with that obtained in \oi.

%
%
%
%
%
%

\section{\texttt{cloudy} implemantation of the burstiness}\label{appendix:bursty}

We demonstrate the effect of bursty star formation in \oiii, \cii, and \oi\ line strength using \texttt{cloudy}. 
The fiducial calculations are basically the same as Section \ref{subsec:OICIIOIII}; we introduce three variables of a gas density ($n_{\rm H}$), gas metallicity ($Z$), and ionization parameter ($U_{\rm ion}$) with constant-pressure, Orion-type grains, and a stopping criterion of $A_V=100$. 
Then two classes of stellar populations were considered:
(i) continuous star formation with an age of 100\,Myr, representing a quasi-steady star-forming activity.
(ii) instantaneous burst with ages of $t_{\rm burst}1$, 3, and 10\,Myr.
To mimic bursty star formation histories, we combined (i) and (ii). 
The composite line luminosities were approximated by linearly combining the \texttt{cloudy} outputs from the burst and continuous models with different burst fractions $f_{\rm burst}$ in $L_{\rm total}=L_{\rm cont}+f_{\rm burst}\times L_{\rm burst}(t_{\rm burst})$, where we adopt $f_{\rm burst}=0.1$, 1.0, 10.0.

\begin{figure*}[htbp]
\begin{center}
\epsscale{1.15}
\includegraphics[width=18cm,bb=0 0 1000 650, trim=0 1 0 0cm]{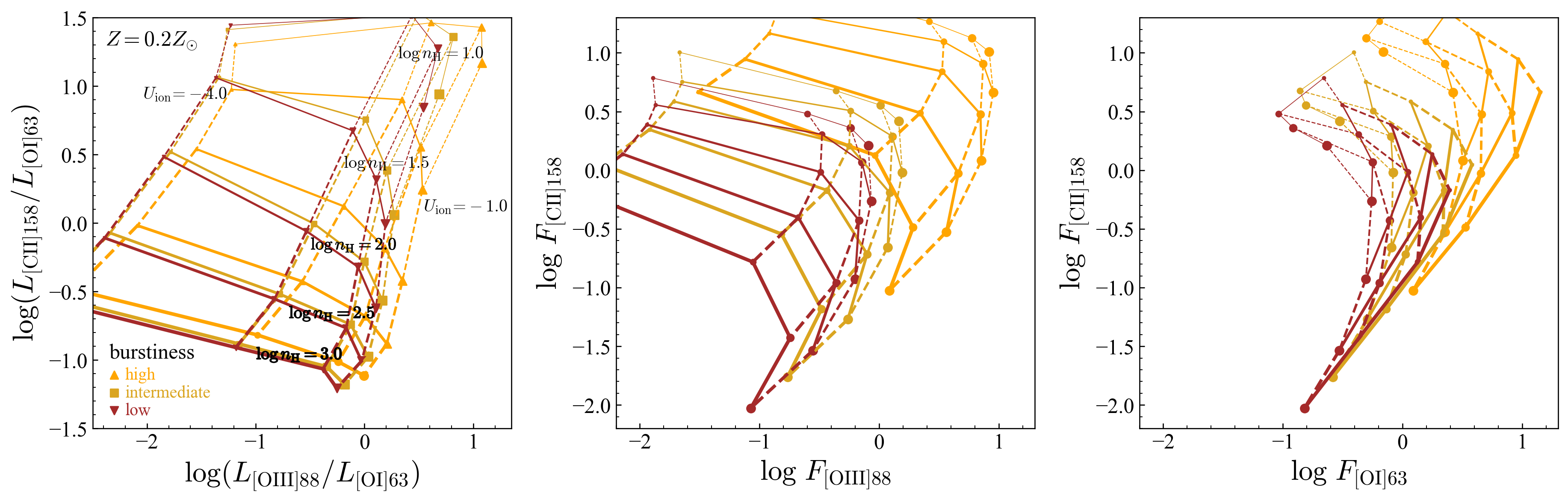}
\caption{Comparison of the \texttt{cloudy} calculation on line ratio between \cii/\oi\ and \oiii/\oi.\ (left), the flux comparison between \cii\ and \oiii\ (middle), and the flux comparison between \cii\ and \oi\ (right) for the bursty star formation effect.
To make the effect of the burstiness easier to understand, we fix $Z=0.2\,Z_{\odot}$ and show a range of $\log n_H=[0.5,3.0]$, $\log U_{\rm ion}=[-4.0,-1.0]$.
For the burstiness models, we show three models with ($t_{\rm burst}$, $f_{\rm burst})=(1\,{\rm Myr}$, 10.0), (3\,Myr, 1.0), and (10\,Myr, 0.1) as a representative high, intermediate, and low burstiness models in the orange, tirangles, brown squares, and dark brown triangles.}
\label{fig:cloudy_burst}
\end{center}
\end{figure*}

The results are shown in Figure \ref{fig:cloudy_burst}.
Here we fix the metallicity to $Z=0.2\,Z_{\odot}$ for simplicity, while there is no major dependence on the metallicity.
For the comparison, we pick three models with ($t_{\rm burst}$, $f_{\rm burst})=(1\,{\rm Myr}$, 10.0), (3\,Myr, 1.0), and (10\,Myr, 0.1), corresponding to ${\rm SFR}_{\rm 10Myr}/{\rm SFR}_{\rm 100Myr}\sim10$, 2, and 1, as high, intermediate, and low burstiness models.
The increase in burstiness changes the stellar SED shape toward a younger, massive star-dominated population, simultaneously enhancing these emission lines in order of \oiii, \oi, and \cii\ (see Section \ref{subsec:highOIIICII} for more discussions about the \texttt{cloudy} results).

%
%
%
%
%
%
%
%
%
%
%
%


\clearpage
\bibliography{main.bib}{}

\bibliographystyle{apj.bst}

\end{document}